 \documentclass{aastex6}

\fullcollaborationName{The Friends of AASTeX Collaboration}
\shorttitle{}
\shortauthors{Katsuda et al.}
\begin{document}

%% LaTeX will automatically break titles if they run longer than
%% one line. However, you may use \\ to force a line break if
%% you desire.

\title{Progenitor Mass Distribution of Core-Collapse Supernova Remnants in Our Galaxy and Magellanic Clouds based on Elemental Abundances}

%% Use \author, \affil, plus the \and command to format author and affiliation 
%% information.  If done correctly the peer review system will be able to
%% automatically put the author and affiliation information from the manuscript
%% and save the corresponding author the trouble of entering it by hand.
%%
%% The \affil should be used to document primary affiliations and the
%% \altaffil should be used for secondary affiliations, titles, or email.

%% Authors with the same affiliation can be grouped in a single
%% \author and \affil call.
\author{Satoru Katsuda\altaffilmark{1}, Tomoya Takiwaki\altaffilmark{2}, 
Nozomu Tominaga\altaffilmark{3,4}, Takashi J.\ Moriya\altaffilmark{2}, 
and Ko Nakamura\altaffilmark{5}
}

\altaffiltext{1}{Graduate School of Science and Engineering, Saitama University, 255 Shimo-Ohkubo, Sakura, Saitama 338-8570, Japan; katsuda@phy.saitama-u.ac.jp}

\altaffiltext{2}{National Astronomical Observatory of Japan, Osawa, Mitaka, Tokyo 181-8588, Japan}

\altaffiltext{3}{Department of Physics, Faculty of Science and Engineering, Konan University, 8-9-1 Okamoto, Kobe, Hyogo 658-8501, Japan}

\altaffiltext{4}{Kavli Institute for the Physics and Mathematics of the Universe (WPI), The University of Tokyo, 5-1-5 Kashiwanoha, Kashiwa, Chiba 277-8583, Japan}

\altaffiltext{5}{Department of Applied Physics, Fukuoka University, Nanakuma 8-19-1, Jonan, Fukuoka 814-0180, Japan}

\begin{abstract}
We investigate a progenitor mass distribution of core-collapse supernova remnants (CCSNRs) in our Galaxy and the Large and Small Magellanic Clouds, for the first time.  We count the number of CCSNRs in three mass ranges divided by the zero-age main-sequence mass, $M_{\rm ZAMS}$; 
A: $M_{\rm ZAMS} < 15\ {\rm M}_\odot$,
B: $15\ {\rm M}_\odot < M_{\rm ZAMS} < 22.5\ {\rm M}_\odot$,
C: $M_{\rm ZAMS} > 22.5\ {\rm M}_\odot$.
Simple compilation of progenitor masses in the literature yields a progenitor mass distribution of $f_{\rm A}: f_{\rm B}: f_{\rm C} =0.24:0.28:0.48$, where $f$ is the number fraction of the progenitors. The distribution is inconsistent with any standard initial mass functions.  We notice, however, that previous mass estimates are subject to large systematic uncertainties because most of the relative abundances (X/Si) are not really good probe for the progenitor masses.  Instead, we propose to rely only on the Fe/Si ratio which is sensitive to the CO core mass ($M_{\rm COcore}$) and $M_{\rm ZAMS}$.  Comparing Fe/Si ratios in SNRs in the literature with the newest theoretical model, we estimate 33 $M_{\rm COcore}$ and $M_{\rm ZAMS}$, leading to a revised progenitor mass distribution of $f_{\rm A}: f_{\rm B}: f_{\rm C} = 0.47: 0.32 : 0.21$.  This is consistent with the standard Salpeter initial mass function.  However, the relation between $M_{\rm COcore}$ and $M_{\rm ZAMS}$ could be affected by binary evolution, which is not taken into account in this study and should be considered in the future work to derive a better progenitor mass distribution estimate.
\end{abstract}


\section{Introduction} \label{sec:intro}
A fundamental question about evolution of massive stars is in which mass range they explode as core-collapse supernovae (CCSNe) and otherwise collapse to black holes without explosions.  A rough prediction is obtained by stellar evolution theory; the mass range of supernovae is from $\sim 8\ {\rm M}_\odot$ to $\sim 40\ {\rm M}_\odot$ or limitless depending on the metallicity of the progenitors \citep[e.g., Fig.~2 of ][]{2003ApJ...591..288H}.  Although great efforts have been devoted to perform hydrodynamic simulations of collapsing massive stars in multi-dimensions including sophisticated neutrino transport, the simulations have not drawn a robust conclusion on the range \citep{2018SSRv..214...33B,2016ARNPS..66..341J,2012PTEP.2012aA301K}.  Observational clues that connect properties of progenitors, supernovae, and supernova remnants, should play an important role in constraining the theory. 

The information of the progenitors of nearby CCSNe can be obtained by the identification of massive stars which have disappeared after explosions in archival images. Their zero-age main-sequence (ZAMS) mass is determined by comparing the luminosity and the color of the massive star with those of theoretical stellar models. By compiling 8 estimates and 12 upper limits of progenitor masses for Type IIP SNe, \citet{2009MNRAS.395.1409S} noticed that the minimum and maximum masses are 8.5$^{+1.0}_{-1.5}$\,M$_\odot$ and 16.5$\pm$1.5\,M$_\odot$, respectively, assuming a power-law slope $\alpha = -2.35$ expected in the Salpeter initial mass function (IMF) \citep[][]{1955ApJ...121..161S}.  Although the minimum mass is consistent with the expectation, the maximum mass is significantly smaller than the most massive red supergiants (25--30\,M$_\odot$) that are found in Milky Way and Magellanic Clouds; we convert luminosities of the brightest red supergiants \citep{2018MNRAS.tmp.1254D} to the initial masses by using the luminosity-mass relation in \citet{2016MNRAS.463.1269B}. These results suggest that the red supergiants with masses about 16--30\,M$_\odot$ may not explode as Type IIP supernovae, which is known as `red supergiant problem'.  A later study including 18 detections and 27 upper limits of Type IIP supernova progenitors came to the same conclusion that the observed populations of SNe IIP in the local Universe are not produced by high-mass (M$_{\rm ZAMS}\gtrsim$18\,M$_\odot$) stars \citep{2015PASA...32...16S}.  From a theoretical point of view, \citet{2014MNRAS.445L..99H} argued that the massive stars' explodability depends on a compactness parameter $\xi$ \citep{2011ApJ...730...70O}, and that stars in a mass range of 16--30\,M$_\odot$ populate an island of stars with high $\xi$ values and do not produce canonical CCSNe by using data of  two- and three-dimensional CCSN simulations \citep{2015PASJ...67..107N,2016MNRAS.461L.112T}.  This is roughly consistent with the systematic one-dimensional simulations \citep[e.g.,][]{2016ApJ...821...38S,2016ApJ...818..124E,2016ApJ...821...69E,2012ApJ...757...69U,2018arXiv180403182E}.

The data given by the direct identification of supernova progenitors must be interpreted with caution for several reasons.  First, the serendipitous pre-explosion images are still rare; only 30 progenitors are detected at present \citep{2017RSPTA.37560277V}.  In a finite sample, the upper mass limit inferred is strongly dependent on the most massive object in the sample and one will always systematically underestimate the upper mass cutoff.  The systematic error for the current sample size was estimated to be 2\,M$_\odot$ \citep{2018MNRAS.474.2116D}.  Second, mass estimates could be systematically lower at the high-mass end.  \citet{2012MNRAS.419.2054W}, \citet{2016MNRAS.463.1269B}, and \citet{2018MNRAS.474.2116D} suggested that an extra extinction due to circumstellar dusts in stellar winds could resolve the red supergiant problem (but see also \citealt{2012ApJ...759...20K}).  Moreover, progenitors of SNe~Ibc are not included in this study, even though they are thought to be the most massive Wolf-Rayet (WR) stars for a single star evolution.  It is true that there is only one detected WR star \citep{2013ApJ...775L...7C,2013AA...558L...1G} and 14 upper limits for nearby SNe~Ibc \citep{2015PASA...32...16S}, but this fact does not strongly support the lack of high-mass progenitors of SNe~Ibc.  This is because WR stars just before explosions could be too faint to detect in optical bands \citep{2012AA...544L..11Y}.  Therefore, adding SNe~Ibc to the SNe IIP sample could also potentially push up the cutoff of the progenitor mass distribution.

Analysis of stellar population surrounding SNe and SN remnants (SNRs) also provides mass distributions of massive stars.  In this method, one first estimates the age distribution of the local stellar population around cataloged SNRs, and use the distribution to measure the star-formation history of the region.  Then, the peak age in the star-formation history is assumed to be the age of the progenitor, and the age is converted to the mass of the progenitor.  \citet{2009ApJ...700..727B} were the first to determine progenitor masses from resolved stellar populations around CCSNRs in the Large Magellanic Cloud.  \citet{2012ApJ...761...26J,2014ApJ...795..170J} applied this technique to 115 SNRs in M31 and M33.  By fitting the mass distribution with a power-law form $dN/dM \propto M^{\alpha}$, they found a power-law slope $\alpha = -4.2\pm0.3$.  This is significantly steeper than $\alpha = -2.35$ for the Salpeter IMF, suggesting either a paucity of massive stars compared with the Salpeter IMF or a selection bias against SNRs in the youngest ($<$10\,Myr old) star-formation regions.  More recently, using new star-formation histories for 62 SNRs in M33, \citet{2018arXiv180207870D} inferred progenitor masses for 94 SNRs in M31 and M33.  They found the best-fit progenitor mass distribution with a minimum mass of 7.33$^{+0.02}_{-0.16}$\,M$_\odot$, the slope of $\alpha = -2.96^{+0.45}_{-0.25}$, and the maximum mass of $>$59\,M$_\odot$.  This is consistent with the previous results, but came closer to the Salpeter IMF.  \citet{2018arXiv180308112W} constrained the progenitor masses of 12 historic CCSNe and 2 SN imposters, by age-dating the stellar populations.  By combining the new measurements with previous results, they found that the distribution of 25 CCSNe progenitor masses is consistent with the Salpeter IMF.  In addition, \citet{2018arXiv180410210A} measured progenitor mass distribution of CCSNRs in the Small Magellanic Clouds using the same approach, finding that the distribution is similar to the Salpeter IMF.  Under these circumstances, further investigations of progenitor mass distributions have been desired.

In this paper, we present, for the first time, a progenitor mass distribution for CCSNRs in our Galaxy and Large and Small Magellanic Clouds, by compiling individual progenitor masses based on elemental abundances in the literature.  Our sample consists of relatively young and X-ray bright SNRs.  They are subject to different selection biases from those of the other methods used to infer progenitor masses.  For example, our sample in principle includes all types of SNe (IIP, Ibc, etc.) and their progenitors (red supergiant stars, WR stars, etc.), which is in contrast to the sample of direct optical imaging of nearby extragalactic progenitors that would not be sensitive to WR stars.  Interestingly, we find that the mass distribution is consistent with the standard Salpeter IMF, which indicates no evidence for the deficit of the supernova explosions of the most massive stars.  This paper is organized as follows.  In Section 2, we compile the progenitor masses from the literature.  We also point out a problem in the estimate of progenitor masses from elemental abundances, and update the values based on a better way we propose.  In Section 3, we discuss possible systematic uncertainties on the estimates of progenitor masses.  In Section 4, we discuss possible effects of selection biases.  In Section 5, we give conclusions and point out a future prospect in the era of X-ray micro-calorimeters.  

\section{Progenitor mass estimates of Galactic CCSNRs}

\subsection{Previous estimates}

There are mainly two ways to estimate progenitor masses of CCSNRs in our Galaxy and Magellanic Clouds: (1) comparing elemental abundances of SN ejecta with those reproduced in SN nucleosynthesis models \citep[e.g.,][]{2012AARv..20...49V}, (2) using a relation between a size of a stellar-wind bubble and a stellar mass \citep{1999ApJ...511..798C,2013ApJ...769L..16C}.  Here we concentrate on the first method, which is more commonly used and probably more reliable to infer progenitor masses than the second method.  This way we tried to eliminate potential systematic uncertainties on the progenitor mass estimates.  

X-ray emission is generally the most powerful tool to derive elemental abundances of SNRs, given that SNRs are hot enough (10$^7$\,K) to emit X-rays efficiently.  Modern X-ray CCDs onboard {\it Suzaku}, {\it XMM-Newton}, and {\it Chandra} are capable of resolving individual lines from all major $\alpha$ elements (C, O, Ne, Mg, Si, S, Ar, Ca, and Fe), and hence we can measure their relative abundances from X-ray spectra.  It is common that the abundances are given relative to Si, as its K-shell line emission is often one of the strongest lines.  When sufficient photon statistics are available, this method has been routinely applied, resulting in a few tens of progenitor mass estimates for CCSNRs in our Galaxy thus far.  

We list progenitor mass estimates of SNRs in our Galaxy and Magellanic Clouds from the literature in the third column of Table~\ref{tab:snrs}.  The criteria to select these SNRs are either (1) progenitor masses are inferred based on elemental abundances measured in X-rays, ultraviolet, or optical bands, (2) preceding X-ray spectroscopy provided Fe and Si abundances with indications of SN ejecta (see below for this reason), or both of them.  Note that SNR~1987A and the Vela SNR are excluded from our sample, since SNR~1987A is too young to show evidence for SN ejecta in X-ray emission and the Vela SNR is too large for any modern X-ray observatories to cover the whole remnant.  We also do not include some SNRs whose origins (i.e., Type Ia or CC) are still debated.  

Since uncertainties on these progenitor masses are generally large and qualitatively determined in the literature, we categorized them into three mass regions (A) M$_{\rm ZAMS} < 15$\,M$_\odot$, (B) 15\,M$_\odot <$ M$_{\rm ZAMS} < 22.5$\,M$_\odot$, and (C) M$_{\rm ZAMS}$ $>$ 22.5\,M$_\odot$ to be 7$^{+1}_{-5}$, 8$^{+6}_{-5}$, and 14$^{+4}_{-2}$, respectively.  The uncertainties are due to numbers of progenitors with marginal masses.  The numbers of progenitors are then converted to the number fractions as in Table~\ref{tab:frac}. By a naive interpretation, it suggests a top-heavy IMF that is not expected in the environment of solar metallicity.

For comparison, we also computed the fraction for other IMFs with three distinct parameters: (the distribution power-law slope $\alpha$, the minimum mass M$_{\rm min}$, the maximum mass M$_{\rm max}$) = ($-2.35$, 8\,M$_\odot$, 40\,M$_\odot$), ($-3.0$, 8\,M$_\odot$, 40\,M$_\odot$), and ($-2.35$, 8\,M$_\odot$, 16.5\,M$_\odot$), which are responsible for the Salpeter IMF, the progenitor mass disribution for CCSNRs in M31 and M33 \citep{2012ApJ...761...26J,2014ApJ...795..170J,2018arXiv180207870D}, and that for SNe IIP \citep{2009MNRAS.395.1409S}, respectively.  These are in stark contrast to that obtained for our sample that are simply taken from the literature.

\subsection{Updating abundance-based progenitor masses}\label{sec:updatedmethod}

\begin{figure}[ht!]
%\figurenum{2}
\gridline{\fig{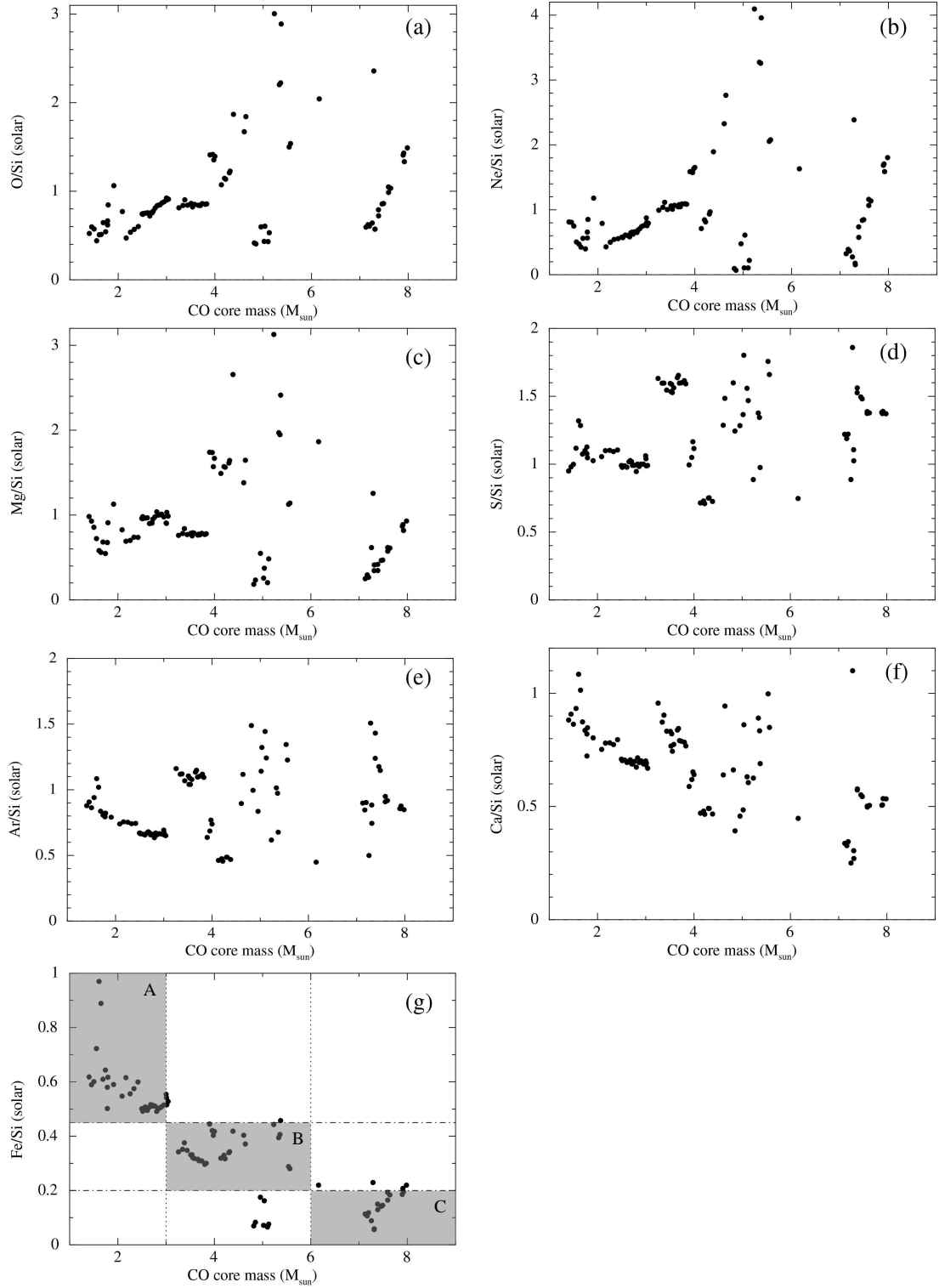}{0.85\textwidth}{}
          }
\caption{(a)--(g) Abundance ratios (X/Si) as a function of the CO core mass, based on nucleosynthesis models by \citet{2016ApJ...821...38S}.  The abundances are relative to the solar values \citep{1989GeCoA..53..197A}.  In the panel (g), dotted and dash-dotted lines indicate the progenitor-mass classes: A, B, and C for  M$_{\rm COcore} <$ 3\,M$_\odot$ $\Leftrightarrow$ M$_{\rm ZAMS} <$ 15\,M$_\odot$, 3\,M$_\odot$ $<$ M$_{\rm COcore} <$ 6\,M$_\odot$ $\Leftrightarrow$ 15\,M$_\odot$ $<$ M$_{\rm ZAMS} < $22.5\,M$_\odot$, and 6\,M$_\odot$ $<$ M$_{\rm COcore}$ $\Leftrightarrow$ 22.5\,M$_\odot$ $<$ M$_{\rm ZAMS}$, respectively.  
\label{fig:abund}}
\end{figure}

We realized that previous estimates of the progenitor masses based on the elemental abundances are not appropriate in many cases as described below.  The masses have been inferred by comparing observed relative abundances (X/Si) with those in the nucleosynthesis models having various ZAMS masses (M$_{\rm ZAMS}$).  When matching, all the relative abundances are equally weighted (usually without considering the statistical errors).  However, most of the X/Si ratios are in fact never good indicator of the progenitor masses.  This can be recognized in Fig.~\ref{fig:abund}, where we plot X/Si for 95 nucleosynthesis models with different M$_{\rm ZAMS}$ that are taken from the state-of-the-art numerical simulations in this context \citep{2016ApJ...821...38S}.  We note that x-axises are taken as a progenitor's CO core mass (M$_{\rm COcore}$) as a proxy of M$_{\rm ZAMS}$ since M$_{\rm COcore}$ is directly related to the nucleosynthesis yields of elements heavier than C and/or the explosion mechanism.  Later we will show how to convert the CO core mass to the ZAMS mass. 
Only Fe/Si and Ca/Si ratios show moderate dependency on the core masses; the correlation coefficients for O/Si, Ne/Si, Mg/Si, S/Si, Ar/Si, Ca/Si, and Fe/Si are calculated to be $0.32, 0.22, -0.11, 0.32, 0.26, -0.67, -0.84$, respectively.  In other words, the Fe/Si and Ca/Si abundance ratios can be good measures of M$_{\rm COcore}$ (rather than M$_{\rm ZAMS}$) regardless of a star's evolutionary path because of the short evolutionary timescale (a few kyr or less) after the carbon burning.  In addition, we checked that the other nucleosynthesis model by \citet{1995ApJS..101..181W} also shows a clear anti-correlation between Fe/Si and M$_{\rm ZAMS}$, which is qualitatively consistent with that in \citet{2016ApJ...821...38S}.

Given the above consideration, we decided to reassess the progenitor masses (M$_{\rm COcore}$ and M$_{\rm ZAMS}$), based solely on the Fe/Si ratio.  We disregard the Ca/Si ratio because it is usually constrained much more loosely than the Fe/Si ratio in observations.  In the third column of Table~\ref{tab:snrs}, we list the Fe/Si abundance ratios taken from the literature.  In some cases, no constraints on Fe abundances were derived, for which we leave the Fe/Si column blank (``not available" or N.A.\ for short).  Since the Fe/Si--M$_{\rm COcore}$ correlation is moderate (see panel (g) of Fig.~\ref{fig:abund}), we do not determine M$_{\rm COcore}$ values for individual SNRs, but rather categorized them into three classes: (A) M$_{\rm COcore} <$ 3\,M$_\odot$, (B) 3\,M$_\odot$ $<$ M$_{\rm COcore} <$ 6\,M$_\odot$, and (C) 6\,M$_\odot$ $<$ M$_{\rm COcore}$.  Two thresholds, (Fe/Si)/(Fe/Si)$_\odot$ $=$ 0.45 and 0.2, can reasonably divide the three classes, as shown in panel (g) of Fig.~\ref{fig:abund}.  The solar abundance is based on \cite{1989GeCoA..53..197A}, since this abundance table is commonly used in the literature listed in Table~\ref{tab:snrs}.  

There is a cluster of outliers in the class B at $M_{\rm COcore} \sim 5$\,M$_{\odot}$ or $M_{\rm ZAMS} \sim 20$\,M$_{\odot}$.  In that mass range, the compactness parameter shows a large scatter, $ 0.1 < \xi_{2.5} < 0.3 $, due to complicated coupling of convection and nucleosynthesis during the stellar evolution (see Fig.~1 of \citealt{2016ApJ...821...38S}).  The models with $\xi_{2.5} \sim 0.1$  in the mass range correspond to the outliers, whereas the other models with $\xi_{2.5} \ge 0.2$ are categorized as the class B.  Due to the small $\xi_{2.5}$, the central engines of the outlier models are similar to that of the lighter stars ($M_{\rm ZAMS} < 12M_{\odot}$).  Since the explosion energies of these outliers are relatively small, the ejected amount of $^{56}$Ni is not large.  However, as mentioned by \citet{2016ApJ...821...38S} and \citet{2016ApJ...818..124E}, we can not ignore a contribution from a neutrino-driven wind that is not sufficiently considered in the explosion models.  In the neutrino-driven wind, iron-group trace elements are synthesized and some of them contribute to $^{56}$Ni.  Unfortunately, it is difficult to precisely estimate its contribution because the nucleosynthesis in the neutrino-driven wind strongly depends on their dynamics and neutrino transport.  However, there is an argument that the actual $^{56}$Ni synthesis is between M($^{56}$Ni) and M(trace elements), and the true Ni masses for these progenitors could be larger by a factor of two, making them typical as the class B.  While this correction should be applied for all the models of smaller compactness parameter in principle, the impact of this correction to our study is the most significant for the outliers.

\begin{figure}[ht!]
%\figurenum{2}
\gridline{\fig{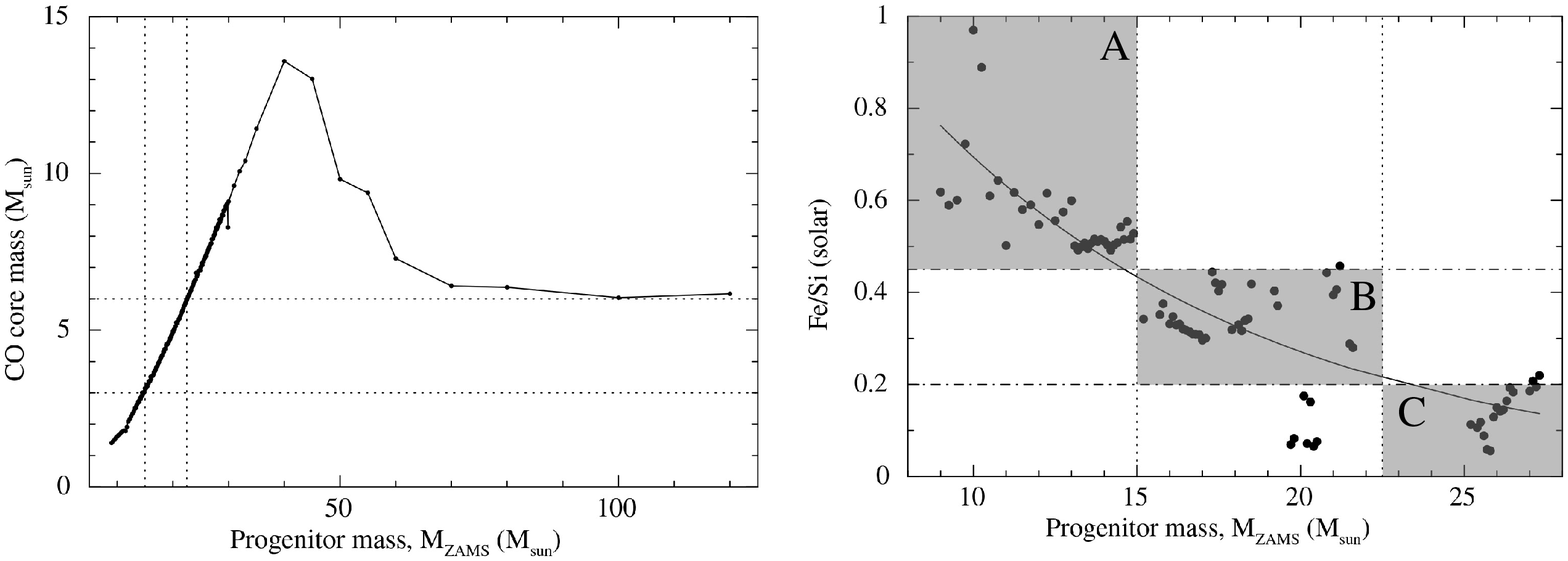}{0.9\textwidth}{}
          }
\caption{{\it Left}: CO core mass as a function of a progenitor mass, M$_{\rm ZAMS}$ \citep{2016ApJ...821...38S}.  The vertical and horizontal dotted lines indicate ZAMS masses of 15\,M$_\odot$ and 22.5\,M$_\odot$ corresponding to CO core masses of 3\,M$_\odot$ and 6\,M$_\odot$, respectively.  {\it Right}: Fe/Si abundance ratios as a function of the progenitor mass (M$_{\rm ZAMS}$), based on nucleosynthesis models by \citet{2016ApJ...821...38S}.  The progenitor classes (A), (B), and (C) are shown in the same manner as in panel (g) of Fig.~\ref{fig:abund}.  The solid curve shows the best-fit exponential model, (Fe/Si)/(Fe/Si)$_\odot$ $=$ 1.13$\times \exp\left(\frac{4.8 - {\rm M}_{\rm ZAMS}}{10.6}\right)$.  
\label{fig:zams}}
\end{figure}

Conversion of M$_{\rm COcore}$ to M$_{\rm ZAMS}$ is also important, since M$_{\rm ZAMS}$ has been used in the literature on progenitor mass distributions.  In fact, the M$_{\rm COcore}$--M$_{\rm ZAMS}$ relation shown in Fig.~\ref{fig:zams} left \citep{2016ApJ...821...38S} assures that the three M$_{\rm COcore}$ classes (A), (B), and (C) given above correspond to M$_{\rm ZAMS} <$ 15\,M$_\odot$, 15\,M$_\odot$ $<$ M$_{\rm ZAMS} <$ 22.5\,M$_\odot$, and M$_{\rm ZAMS}$ $>$ 22.5\,M$_\odot$ for a single star evolution, respectively.  Moreover, there is a good linear relation between M$_{\rm COcore}$ and M$_{\rm ZAMS}$ at a mass range of M$_{\rm ZAMS}$ $\lesssim$ 40\,M$_\odot$ as shown in Fig.~\ref{fig:zams} left.  Together with a good Fe/Si--M\,$_{\rm COcore}$ correlation in panel (g) of Fig.~\ref{fig:abund}, there is also a good Fe/Si--M\,$_{\rm ZAMS}$ correlation, which is shown in Fig.~\ref{fig:zams} right.  The data points can be fitted by an exponential function, (Fe/Si)/(Fe/Si)$_\odot$ $=$ 1.13$\times \exp\left(\frac{4.8 - {\rm M}_{\rm ZAMS}}{10.6}\right)$, which is shown as a solid curve in the right panel of Fig.~\ref{fig:zams}.

\begin{figure}[ht!]
%\figurenum{2}
\gridline{\fig{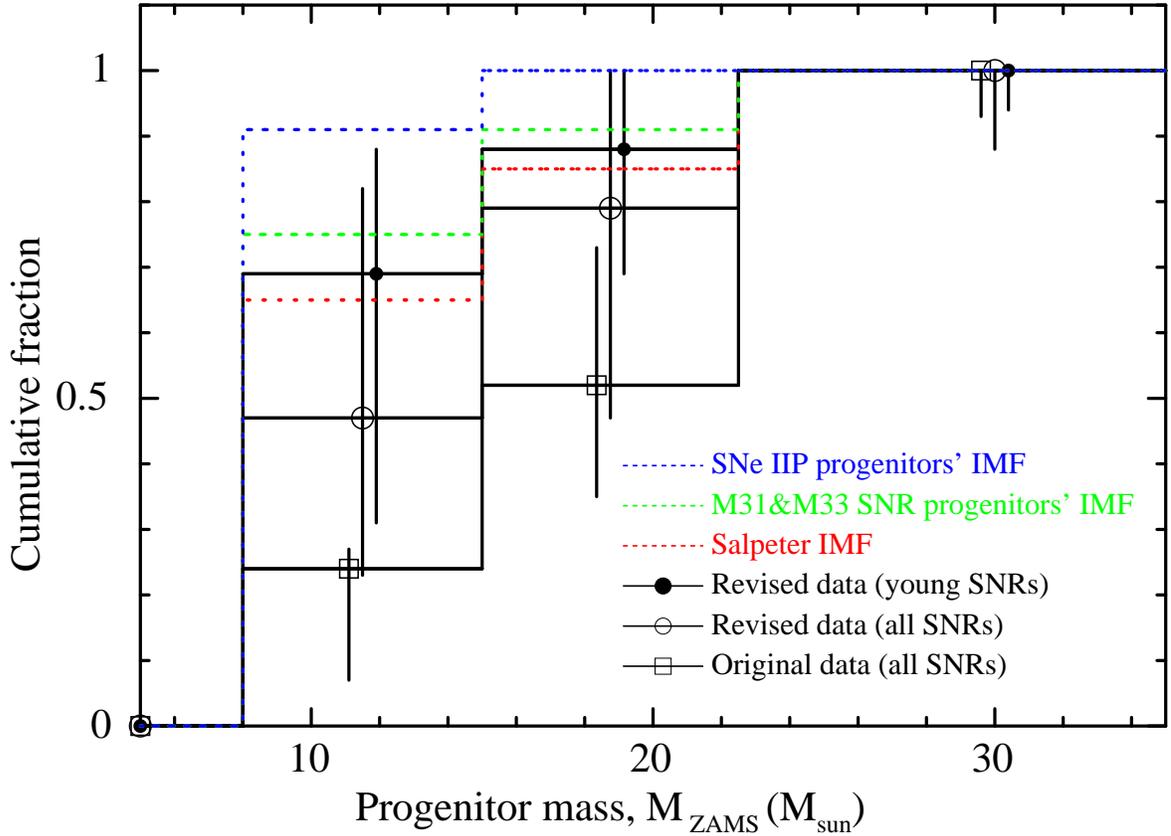}{0.9\textwidth}{}
          }
\caption{Cumulative distributions of progenitor masses (M$_{\rm ZAMS}$) for six different cases.  The black lines with open box, open circle, and filled circle are responsible for Galactic and Magellanic Clouds SNRs that are simply taken from the literature (the third column in Table~\ref{tab:snrs}), revised based on Fe/Si ratios (the sixth column in Table~\ref{tab:snrs}), and the revised data restricted to the young ($t < 5000$\,yr) SNRs, respectively.  The dashed lines in red, green, and blue are responsible for the standard Salpeter IMF, the up-to-date progenitors' IMF for CCSNRs in M31 and M33 \citep{2018arXiv180207870D}, and progenitors' IMF for SNe IIP \citep{2009MNRAS.395.1409S}.
\label{fig:cum_frac}}
\end{figure}

The revised classifications of progenitor masses are summarized in the fourth and fifth columns of Table~\ref{tab:snrs}.  We then re-calculated the number of progenitors to be 13$^{+10}_{-7}$, 10$^{+11}_{-10}$, 7$^{+9}_{-4}$ for classes (A), (B), and (C), respectively.  Given that younger SNRs are often more sensitive to SN ejecta than older SNRs, we also calculated the number of progenitors for relatively young SNRs with ages less than $\sim$5000\,yr to be 11$^{+7}_{-5}$, 4$^{+6}_{-4}$, 2$^{+3}_{-1}$ for classes (A), (B), and (C), respectively.  Note that the errors are due either to ``marginal" progenitors between the boundaries or to different estimates from different methods.  The ``marginal" progenitors are identified when the observed Fe/Si ratios are consistent with one of the two boundaries with a potential uncertainty of 0.1 dex on the model: 0.15--0.25 and 0.33--0.57.  These marginal data are indicated by m1 or m2 in Table~\ref{tab:snrs}, where m1 and m2 correspond to the lower- and higher-mass boundaries, respectively.  We give the number fractions in Table~\ref{tab:frac}, and plot their cumulative distributions in Fig.~\ref{fig:cum_frac} for a demonstration.

Notably, most of the revised progenitor masses are significantly different from the previous estimates, suggesting the importance to focus on the Fe/Si abundance ratio rather than taking the balance of the overall abundance pattern.  In addition, the revised progenitor mass distribution is found to be consistent with the standard Salpeter IMF.  If single stars are major contributors to CCSNe, then our result suggests that there is no high-mass cutoff for progenitors of CCSNe, which conflicts to the deficit of the supernova explosions of the most massive stars based on the previous works on the age-dating of a stellar population around CCSNRs in M31 and M33 \citep[e.g.,][]{2014ApJ...795..170J}. The direct imaging of SN progenitors \citep[e.g.,][]{2009MNRAS.395.1409S} also suggests the high-mass cutoff in the distribution.  It is, however,  not easy to compare our result with that of the direct imaging since that does not  include the contribution of SNe Ibc, which is included in our analysis.

Our findings are supported by several studies. From a theoretical point of view, heavier progenitors tend to explode relatively easily despite the strong mass accretion onto a proto-neutron star in the recent simulations that  include general relativistic effects and employ  newer progenitor sets \citep{2018ApJ...854...63O,2018ApJ...855L...3O,2018MNRAS.tmp..786V,2016ApJ...825....6S}.  From a view point of chemical evolution, CCSNe should occur in a mass range of 9--100$\ {\rm M}_{\odot}$ to explain oxygen enrichment of our Galaxy \citep{2013ApJ...769...99B,2018ApJ...852..101S}.

As mentioned above, it should be noted that the Fe/Si abundance ratios allow us to directly estimate M$_{\rm COcore}$ rather than M$_{\rm ZAMS}$.  Therefore, it is more robust to argue that there is no high-mass cutoff for CO core masses of progenitor stars.  This argument holds even if effects of binary stars are important.  

\section{Possible uncertainties on the progenitor mass estimates}

We have compiled and updated the progenitor masses for Galactic and Magellanic Clouds CCSNRs in Section 2.  Here we discuss their systematic uncertainties.  First, progenitor masses based on elemental abundances suffer from the observational uncertainties including (1) potential presence of unshocked (invisible) SN ejecta, (2) imperfect elimination of the swept-up medium in X-ray spectra, (3) possible contamination of Type Ia SNRs, and (4) lack of knowledge on emissivities especially for Fe L lines.  

The first point, if essential, should reduce the estimated progenitor masses, because unshocked ejecta are likely rich in Fe and tend to increase the Fe/Si ratio than the value observed.  This effect is certainly important for young SNRs that retain ejecta stratification.  However, the dynamical evolution of CCSNRs tend to be faster than that of Type Ia SNRs due to interactions with the circumstellar medium \citep{2011ApJ...732..114L,2014ApJ...785L..27Y}.  In addition, even in the youngest SNR in our sample, Cas~A, there is an argument that $\sim$90\% of the total ejecta were already heated by a reverse shock \citep{2012ApJ...746..130H,2014ApJ...785....7D}.  Therefore, the effect of unshocked ejecta might not play an important role for CCSNRs.  Still, we cannot exclude the possibility that Cas~A is a special case, and other older CCSNRs contain more fractions of unshocked ejecta than Cas~A (in fact, we suspect that RX~J1713.7-3946 would be such an example \citep{2015ApJ...814...29K}).  In this case, we would under/over-estimate the number of low/high-mass stars.

The second point should result in misidentification of the most massive stars as less massive stars, since the Fe/Si abundance of the swept-up medium is likely close to the solar value and is consistent with the progenitors with $M_{\rm ZAMS} < 15$\,M$_\odot$.  The same effect is expected for the third point, because nucleosynthesis models for SNe Ia predict high Fe/Si ratios \citep[e.g., (Fe/Si)/(Fe/Si)$_\odot$ $\sim$ 1.8 for the W7 model:][]{1984ApJ...286..644N}, and thus some of the lower-mass stars in Table~\ref{tab:snrs} might originate from SNe~Ia.  The last point could alter the observed Fe/Si ratio in either direction \citep{2012ApJ...756..128F}, hence it is difficult to assess its impact on the progenitor mass distribution.  In summary, these systematic effects in total would increase the number of the most massive stars, further arguing against the lack of the most massive stars for CCSNe.

Nucleosynthesis models needed for comparison with observations are also subject to some uncertainties.  For example, the Fe/Si ratio depends on the mass-cut \citep[e.g.,][]{2007ApJ...660..516T} and the explosion energy \citep[e.g.,][]{2018ApJ...856...63F}.  If the real mass cuts of our sampled SNRs would be systematically higher than those in \citet{2016ApJ...821...38S}, then the progenitor masses would be estimated systematically lower.  If the explosion energy of our sampled SNRs would be larger than those in \citet{2016ApJ...821...38S}, then the progenitor masses would be estimated systematically higher.  Given that there are no systematic trends of the two parameters for Galactic CCSNRs and that our reference nucleosynthesis models shown in Fig.~\ref{fig:abund} are calculated based on standard values in terms of these parameters \citep{2016ApJ...821...38S}, we expect no systematic over/under estimates for our progenitor mass estimates.  

Another uncertainty is the effect of binary star progenitors.  This effect would be particularly important, because more than 70\% of all massive stars are in binaries \citep{2012Sci...337..444S}.  For example, based on a 1D hydrodynamic simulation, \citet{2007AA...465L..29C} realized that the evolution of a mass gainer in a 16 + 15\,M$_\odot$ binary system is almost identical to that of a 24\,M$_\odot$ single star.  Namely, the nucleosynthesis yields from the mass gainer (the 15~M$_\odot$ star) and the 24\,M$_\odot$ star should be identical with each other. This means that the 15~M$_\odot$ star explodes like a single 24~M$_\odot$ star in terms of the nucleosynthesis yields.  Therefore, the explosion we estimate as a 24~M$_\odot$ progenitor could actually be a 15~M$_\odot$ progenitor and the binary effect could apparently increase the fraction of the most massive stars.  However, the whole effects of binary stars may be more complicated, requiring further detailed theoretical works.

\section{Possible effects of selection biases}

It is also important to discuss potential selection biases present in our sample.  \citet{2017MNRAS.464.2326S} examined selection effects in radio samples of SNRs.  They found that SNRs in regions of dense ambient medium will be brighter but remain visible for shorter periods of time than SNRs in tenuous medium, with the visibility time scaled by the ambient HI column density: $t_{\rm vis} \propto N_{\rm H}^{-0.33}$.  Here we assume that the same argument can be applied to our X-ray selected samples.  Given that there is a linear correlation between a radius of a stellar wind bubble and a progenitor mass \citep{1999ApJ...511..798C,2013ApJ...769L..16C}, it would be reasonable to expect a lower ambient $N_{\rm H}$ around a more massive star than that of a less massive one.  Therefore, we expect longer lifetimes for SNRs originating from more massive stars, although in reality the presence of a circumstellar medium may reduce the lifetimes to some extent.  This would cause a selection bias to massive stars in our sample.  Such a selection bias, if important, should be mitigated by focusing on SNRs much younger than a typical visibility lifetime of SNRs \citep[a few 10$^5$\,yr based on Fig.~8 in][]{2017MNRAS.464.2326S}.  Therefore, in Table~\ref{tab:frac} we examined the progenitor mass distribution for a sample of young SNRs with ages less than $\sim$5000\,yr, two orders of magnitude younger than the visibility lifetime in radio.  Then, we found that the mass distribution comes closer to the Salpeter IMF than that of all SNRs (although both mass distributions are consistent with the Salpeter IMF).  This might suggest the presence of the selection bias on visibility lifetimes of SNRs.

Also, there is growing evidence that progenitors of some CC SNe undergo enhanced or extreme mass loss prior to explosion.  Such a mass loss substantially affects the dynamical evolution of SNRs, and could be another source of selection biases.  Recently, \citet{2017ApJ...849..109P} performed self-consistent end-to-end simulations for the evolution of a massive star from the pre-main sequence, up to and through core-collapse, and into the remnant phase.  They found that the mass loss in late stages (during and after core carbon burning) can have a profound impact on the dynamics and spectral evolution of the SNR centuries after core collapse.  However, they noted that there is little impact on the late-time dynamics when the forward shock breaks through the circumstellar shell (see also \citealt{2015ApJ...803..101P}).  Given that all of our sampled SNRs are older than a few centuries, the effect of the extreme mass loss prior to explosion would not play an important role for our sample.

\section{Conclusions and Future Prospect}

We have examined the progenitor mass distribution for CCSNRs in our Galaxy and Magellanic Clouds for the first time.  By compiling the progenitor masses from the elemental abundances of the SN ejecta in the literature, and updating 33 progenitor masses based on the Fe/Si abundance ratio, we revealed the progenitors' number fractions to be 
$f_A:f_B:f_C$ = 0.47$^{+0.35}_{-0.24}$ : 0.32$\pm$0.32 : 0.21$^{+0.26}_{-0.12}$, where A, B and C are the mass ranges of 
A: M$_{\rm COcore} <$ 3\,M$_\odot$ or $M_{\rm ZAMS} < 15\ {\rm M}_\odot$,
B: 3\,M$_\odot$ $<$ M$_{\rm COcore} <$ 6\,M$_\odot$ or $15\ {\rm M}_\odot < M_{\rm ZAMS} < 22.5\ {\rm M}_\odot$, and 
C: 6\,M$_\odot$ $<$ M$_{\rm COcore}$ or $22.5\ {\rm M}_\odot < M_{\rm ZAMS}$.
The error comes from marginal progenitors (see Sec.~\ref{sec:updatedmethod} for detail).  This is in agreement with the standard Salpeter IMF.  Therefore, if single stars are dominant contributors to CCSNe, we can argue that there is no high-mass cutoff in exploding massive stars.  This result is apparently in tension with the suggestion by the stellar population around CCSNRs in M31 and M33 galaxies, and also seems to disagree with the suggestion by direct imaging of SN progenitors in nearby galaxies with a caveat that our sample (all kinds of progenitors) is not the same as that of direct imaging (basically only red supergiant stars).  We should keep in mind, however, that there is growing evidence that binary stars contribute significantly to CCSNe.  Therefore, binary effects must be carefully taken into account in the future work to deduce a solid conclusion on the progenitor mass distribution of CCSNRs in our Galaxy.  One robust conclusion from our current investigation is that progenitor stars with massive CO cores do explode, which is true even if effects of binary stars are important.  
Another important argument is that, in estimating a progenitor mass by a comparison between observed elemental abundances and nucleosynthesis models, it is essential to focus on the Fe/Si abundance ratio instead of fitting the various relative abundances (X/Si).  This is because most of the X/Si relative abundances are never sensitive to the progenitor masses, but only Fe/Si is.  

\begin{figure}[ht!]
%\figurenum{2}
\gridline{\fig{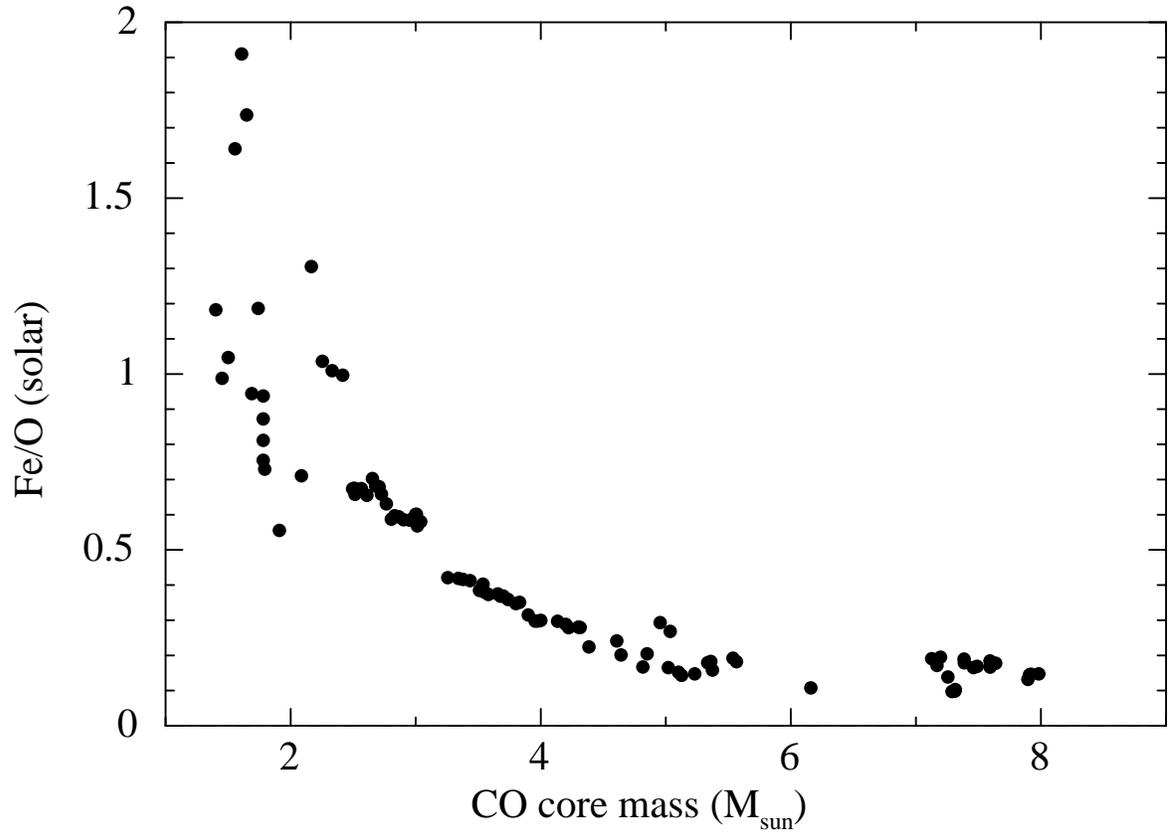}{0.9\textwidth}{}
          }
\caption{Fe/O abundance ratios as a function of the CO core mass, based on nucleosynthesis models by \citet{2016ApJ...821...38S}.  There is a good correlation especially above M$_{\rm COcore} = 2$\,M$_\odot$.
\label{fig:fe2o}}
\end{figure}

Finally, we point out that the Fe/O abundance ratio shown in Fig.~\ref{fig:fe2o} is more sensitive to the progenitor mass than Fe/Si.  This is particularly true for those with M$_{\rm COcore} \gtrsim 2$\,M$_{\odot}$ or M$_{\rm ZAMS} \gtrsim 12$\,M$_{\odot}$.  Therefore, in principle, the Fe/O ratio can work better in estimating the progenitor mass with M$_{\rm ZAMS} \gtrsim 12$\,M$_{\odot}$.  There are two major reasons why we did not use the Fe/O ratio.  First, there are not so many SNRs for which we can measure O abundances owing to severe interstellar absorptions.  Second, even if O lines are detected, they are usually dominated by the swept-up medium rather than SN ejecta.  In this case, one needs to eliminate the swept-up component very carefully in spectral analyses, which is often subject to a large uncertainty.   

We anticipate that the micro-calorimeter onboard X-Ray Imaging and Spectroscopy Mission (XRISM) scheduled to be launched in 2021 will dramatically change this situation.  Its superior energy resolution (FWHM$\sim$5\,eV at 6\,keV) and a good response at the soft X-ray band will substantially increase the number of SNRs with detected O lines.  C and N lines will be also detected from many SNRs, which will ease the separation between the two (swept-up and ejecta) components.  Therefore, the Fe/O ratio will be an important observable for progenitor mass estimations in the era of XRISM.

\acknowledgments

We thank the anonymous referee for his/her constructive comments that helped improve the quality of this paper.  This work was supported by the Japan Society for the Promotion of Science KAKENHI grant numbers JP16K17673, JP17H02864 (SK), JP17K14306, JP17H05206, JP17H06364, JP17H01130 (TT), JP16H07413, JP17H02864, JP18K13585 (TM), JP16K17668, JP17H05205 (KN). KN is also supported by funds from the Central Research Institute of Fukuoka University (Nos. 171042, 177103). This work was partly supported by Leading Initiative for Excellent Young Researchers, MEXT, Japan.

%\bibliography{reference}

\begin{thebibliography}{}
\expandafter\ifx\csname natexlab\endcsname\relax\def\natexlab#1{#1}\fi

\bibitem[{{Anders} \& {Grevesse}(1989)}]{1989GeCoA..53..197A}
{Anders}, E., \& {Grevesse}, N. 1989, \gca, 53, 197

\bibitem[{{Auchettl} {et~al.}(2018){Auchettl}, {Lopez}, {Badenes},
  {Ramirez-Ruiz}, {Beacom}, \& {Holland-Ashford}}]{2018arXiv180410210A}
{Auchettl}, K., {Lopez}, L., {Badenes}, C., {et~al.} 2018, ArXiv e-prints,
  arXiv:1804.10210

\bibitem[{{Badenes} {et~al.}(2009){Badenes}, {Harris}, {Zaritsky}, \&
  {Prieto}}]{2009ApJ...700..727B}
{Badenes}, C., {Harris}, J., {Zaritsky}, D., \& {Prieto}, J.~L. 2009, \apj,
  700, 727

\bibitem[{{Bamba} {et~al.}(2018){Bamba}, {Ohira}, {Yamazaki}, {Sawada},
  {Terada}, {Koyama}, {Miller}, {Yamaguchi}, {Katsuda}, {Nobukawa}, \&
  {Nobukawa}}]{2018ApJ...854...71B}
{Bamba}, A., {Ohira}, Y., {Yamazaki}, R., {et~al.} 2018, \apj, 854, 71

\bibitem[{{Beasor} \& {Davies}(2016)}]{2016MNRAS.463.1269B}
{Beasor}, E.~R., \& {Davies}, B. 2016, \mnras, 463, 1269

\bibitem[{{Becker} {et~al.}(2012){Becker}, {Prinz}, {Winkler}, \&
  {Petre}}]{2012ApJ...755..141B}
{Becker}, W., {Prinz}, T., {Winkler}, P.~F., \& {Petre}, R. 2012, \apj, 755,
  141

\bibitem[{{Blair} {et~al.}(2000){Blair}, {Morse}, {Raymond}, {Kirshner},
  {Hughes}, {Dopita}, {Sutherland}, {Long}, \& {Winkler}}]{2000ApJ...537..667B}
{Blair}, W.~P., {Morse}, J.~A., {Raymond}, J.~C., {et~al.} 2000, \apj, 537, 667

\bibitem[{{Bocchino} {et~al.}(2009){Bocchino}, {Miceli}, \&
  {Troja}}]{2009AA...498..139B}
{Bocchino}, F., {Miceli}, M., \& {Troja}, E. 2009, \aap, 498, 139

\bibitem[{{Broersen} \& {Vink}(2015)}]{2015MNRAS.446.3885B}
{Broersen}, S., \& {Vink}, J. 2015, \mnras, 446, 3885

\bibitem[{{Broersen} {et~al.}(2011){Broersen}, {Vink}, {Kaastra}, \&
  {Raymond}}]{2011AA...535A..11B}
{Broersen}, S., {Vink}, J., {Kaastra}, J., \& {Raymond}, J. 2011, \aap, 535,
  A11

\bibitem[{{Brown} \& {Woosley}(2013)}]{2013ApJ...769...99B}
{Brown}, J.~M., \& {Woosley}, S.~E. 2013, \apj, 769, 99

\bibitem[{{Burrows} {et~al.}(2018){Burrows}, {Vartanyan}, {Dolence}, {Skinner},
  \& {Radice}}]{2018SSRv..214...33B}
{Burrows}, A., {Vartanyan}, D., {Dolence}, J.~C., {Skinner}, M.~A., \&
  {Radice}, D. 2018, \ssr, 214, 33

\bibitem[{{Camilo} {et~al.}(2004){Camilo}, {Gaensler}, {Gotthelf}, {Halpern},
  \& {Manchester}}]{2004ApJ...616.1118C}
{Camilo}, F., {Gaensler}, B.~M., {Gotthelf}, E.~V., {Halpern}, J.~P., \&
  {Manchester}, R.~N. 2004, \apj, 616, 1118

\bibitem[{{Cantiello} {et~al.}(2007){Cantiello}, {Yoon}, {Langer}, \&
  {Livio}}]{2007AA...465L..29C}
{Cantiello}, M., {Yoon}, S.-C., {Langer}, N., \& {Livio}, M. 2007, \aap, 465,
  L29

\bibitem[{{Cao} {et~al.}(2013){Cao}, {Kasliwal}, {Arcavi}, {Horesh}, {Hancock},
  {Valenti}, {Cenko}, {Kulkarni}, {Gal-Yam}, {Gorbikov}, {Ofek}, {Sand},
  {Yaron}, {Graham}, {Silverman}, {Wheeler}, {Marion}, {Walker}, {Mazzali},
  {Howell}, {Li}, {Kong}, {Bloom}, {Nugent}, {Surace}, {Masci}, {Carpenter},
  {Degenaar}, \& {Gelino}}]{2013ApJ...775L...7C}
{Cao}, Y., {Kasliwal}, M.~M., {Arcavi}, I., {et~al.} 2013, \apjl, 775, L7

\bibitem[{{Carter} {et~al.}(1997){Carter}, {Dickel}, \&
  {Bomans}}]{1997PASP..109..990C}
{Carter}, L.~M., {Dickel}, J.~R., \& {Bomans}, D.~J. 1997, \pasp, 109, 990

\bibitem[{{Chen} \& {Slane}(2001)}]{2001ApJ...563..202C}
{Chen}, Y., \& {Slane}, P.~O. 2001, \apj, 563, 202

\bibitem[{{Chen} {et~al.}(2013){Chen}, {Zhou}, \& {Chu}}]{2013ApJ...769L..16C}
{Chen}, Y., {Zhou}, P., \& {Chu}, Y.-H. 2013, \apjl, 769, L16

\bibitem[{{Chevalier}(1999)}]{1999ApJ...511..798C}
{Chevalier}, R.~A. 1999, \apj, 511, 798

\bibitem[{{Davies} \& {Beasor}(2018)}]{2018MNRAS.474.2116D}
{Davies}, B., \& {Beasor}, E.~R. 2018, \mnras, 474, 2116

\bibitem[{{Davies} {et~al.}(2018){Davies}, {Crowther}, \&
  {Beasor}}]{2018MNRAS.tmp.1254D}
{Davies}, B., {Crowther}, P.~A., \& {Beasor}, E.~R. 2018, \mnras,
  arXiv:1804.06417

\bibitem[{{DeLaney} {et~al.}(2014){DeLaney}, {Kassim}, {Rudnick}, \&
  {Perley}}]{2014ApJ...785....7D}
{DeLaney}, T., {Kassim}, N.~E., {Rudnick}, L., \& {Perley}, R.~A. 2014, \apj,
  785, 7

\bibitem[{{D{\'{\i}}az-Rodr{\'{\i}}guez}
  {et~al.}(2018){D{\'{\i}}az-Rodr{\'{\i}}guez}, {Murphy}, {Rubin}, {Dolphin},
  {Williams}, \& {Dalcanton}}]{2018arXiv180207870D}
{D{\'{\i}}az-Rodr{\'{\i}}guez}, M., {Murphy}, J.~W., {Rubin}, D.~A., {et~al.}
  2018, ArXiv e-prints, arXiv:1802.07870

\bibitem[{{Ebinger} {et~al.}(2018){Ebinger}, {Curtis}, {Fr{\"o}hlich},
  {Hempel}, {Perego}, {Liebend{\"o}rfer}, \&
  {Thielemann}}]{2018arXiv180403182E}
{Ebinger}, K., {Curtis}, S., {Fr{\"o}hlich}, C., {et~al.} 2018, ArXiv e-prints,
  arXiv:1804.03182

\bibitem[{{Ertl} {et~al.}(2016{\natexlab{a}}){Ertl}, {Janka}, {Woosley},
  {Sukhbold}, \& {Ugliano}}]{2016ApJ...818..124E}
{Ertl}, T., {Janka}, H.-T., {Woosley}, S.~E., {Sukhbold}, T., \& {Ugliano}, M.
  2016{\natexlab{a}}, \apj, 818, 124

\bibitem[{{Ertl} {et~al.}(2016{\natexlab{b}}){Ertl}, {Ugliano}, {Janka},
  {Marek}, \& {Arcones}}]{2016ApJ...821...69E}
{Ertl}, T., {Ugliano}, M., {Janka}, H.-T., {Marek}, A., \& {Arcones}, A.
  2016{\natexlab{b}}, \apj, 821, 69

\bibitem[{{Fesen} {et~al.}(1997){Fesen}, {Winkler}, {Rathore}, {Downes},
  {Wallace}, \& {Tweedy}}]{1997AJ....113..767F}
{Fesen}, R.~A., {Winkler}, F., {Rathore}, Y., {et~al.} 1997, \aj, 113, 767

\bibitem[{{Fesen} {et~al.}(2006){Fesen}, {Hammell}, {Morse}, {Chevalier},
  {Borkowski}, {Dopita}, {Gerardy}, {Lawrence}, {Raymond}, \& {van den
  Bergh}}]{2006ApJ...645..283F}
{Fesen}, R.~A., {Hammell}, M.~C., {Morse}, J., {et~al.} 2006, \apj, 645, 283

\bibitem[{{Finkelstein} {et~al.}(2006){Finkelstein}, {Morse}, {Green},
  {Linsky}, {Shull}, {Snow}, {Stocke}, {Brownsberger}, {Ebbets}, {Wilkinson},
  {Heap}, {Leitherer}, {Savage}, {Siegmund}, \& {Stern}}]{2006ApJ...641..919F}
{Finkelstein}, S.~L., {Morse}, J.~A., {Green}, J.~C., {et~al.} 2006, \apj, 641,
  919

\bibitem[{{Foster} {et~al.}(2012){Foster}, {Ji}, {Smith}, \&
  {Brickhouse}}]{2012ApJ...756..128F}
{Foster}, A.~R., {Ji}, L., {Smith}, R.~K., \& {Brickhouse}, N.~S. 2012, \apj,
  756, 128

\bibitem[{{France} {et~al.}(2009){France}, {Beasley}, {Keeney}, {Danforth},
  {Froning}, {Green}, \& {Shull}}]{2009ApJ...707L..27F}
{France}, K., {Beasley}, M., {Keeney}, B.~A., {et~al.} 2009, \apjl, 707, L27

\bibitem[{{Frank} {et~al.}(2015){Frank}, {Burrows}, \&
  {Park}}]{2015ApJ...810..113F}
{Frank}, K.~A., {Burrows}, D.~N., \& {Park}, S. 2015, \apj, 810, 113

\bibitem[{{Fryer} {et~al.}(2018){Fryer}, {Andrews}, {Even}, {Heger}, \&
  {Safi-Harb}}]{2018ApJ...856...63F}
{Fryer}, C.~L., {Andrews}, S., {Even}, W., {Heger}, A., \& {Safi-Harb}, S.
  2018, \apj, 856, 63

\bibitem[{{Gaensler} {et~al.}(2008){Gaensler}, {Tanna}, {Slane}, {Brogan},
  {Gelfand}, {McClure-Griffiths}, {Camilo}, {Ng}, \&
  {Miller}}]{2008ApJ...680L..37G}
{Gaensler}, B.~M., {Tanna}, A., {Slane}, P.~O., {et~al.} 2008, \apjl, 680, L37

\bibitem[{{Gelfand} {et~al.}(2013){Gelfand}, {Castro}, {Slane}, {Temim},
  {Hughes}, \& {Rakowski}}]{2013ApJ...777..148G}
{Gelfand}, J.~D., {Castro}, D., {Slane}, P.~O., {et~al.} 2013, \apj, 777, 148

\bibitem[{{G{\"o}k} \& {Sezer}(2012)}]{2012MNRAS.419.1603G}
{G{\"o}k}, F., \& {Sezer}, A. 2012, \mnras, 419, 1603

\bibitem[{{Groh} {et~al.}(2013){Groh}, {Georgy}, \&
  {Ekstr{\"o}m}}]{2013AA...558L...1G}
{Groh}, J.~H., {Georgy}, C., \& {Ekstr{\"o}m}, S. 2013, \aap, 558, L1

\bibitem[{{Haberl} {et~al.}(2012){Haberl}, {Filipovi{\'c}}, {Bozzetto},
  {Crawford}, {Points}, {Pietsch}, {De Horta}, {Tothill}, {Payne}, \&
  {Sasaki}}]{2012AA...543A.154H}
{Haberl}, F., {Filipovi{\'c}}, M.~D., {Bozzetto}, L.~M., {et~al.} 2012, \aap,
  543, A154

\bibitem[{{Heger} {et~al.}(2003){Heger}, {Fryer}, {Woosley}, {Langer}, \&
  {Hartmann}}]{2003ApJ...591..288H}
{Heger}, A., {Fryer}, C.~L., {Woosley}, S.~E., {Langer}, N., \& {Hartmann},
  D.~H. 2003, \apj, 591, 288

\bibitem[{{Hendrick} {et~al.}(2005){Hendrick}, {Reynolds}, \&
  {Borkowski}}]{2005ApJ...622L.117H}
{Hendrick}, S.~P., {Reynolds}, S.~P., \& {Borkowski}, K.~J. 2005, \apjl, 622,
  L117

\bibitem[{{Horiuchi} {et~al.}(2014){Horiuchi}, {Nakamura}, {Takiwaki},
  {Kotake}, \& {Tanaka}}]{2014MNRAS.445L..99H}
{Horiuchi}, S., {Nakamura}, K., {Takiwaki}, T., {Kotake}, K., \& {Tanaka}, M.
  2014, \mnras, 445, L99

\bibitem[{{Hughes} {et~al.}(1998){Hughes}, {Hayashi}, \&
  {Koyama}}]{1998ApJ...505..732H}
{Hughes}, J.~P., {Hayashi}, I., \& {Koyama}, K. 1998, \apj, 505, 732

\bibitem[{{Hwang} \& {Laming}(2012)}]{2012ApJ...746..130H}
{Hwang}, U., \& {Laming}, J.~M. 2012, \apj, 746, 130

\bibitem[{{Hwang} {et~al.}(2008){Hwang}, {Petre}, \&
  {Flanagan}}]{2008ApJ...676..378H}
{Hwang}, U., {Petre}, R., \& {Flanagan}, K.~A. 2008, \apj, 676, 378

\bibitem[{{Janka} {et~al.}(2016){Janka}, {Melson}, \&
  {Summa}}]{2016ARNPS..66..341J}
{Janka}, H.-T., {Melson}, T., \& {Summa}, A. 2016, Annual Review of Nuclear and
  Particle Science, 66, 341

\bibitem[{{Jennings} {et~al.}(2012){Jennings}, {Williams}, {Murphy},
  {Dalcanton}, {Gilbert}, {Dolphin}, {Fouesneau}, \&
  {Weisz}}]{2012ApJ...761...26J}
{Jennings}, Z.~G., {Williams}, B.~F., {Murphy}, J.~W., {et~al.} 2012, \apj,
  761, 26

\bibitem[{{Jennings} {et~al.}(2014){Jennings}, {Williams}, {Murphy},
  {Dalcanton}, {Gilbert}, {Dolphin}, {Weisz}, \&
  {Fouesneau}}]{2014ApJ...795..170J}
---. 2014, \apj, 795, 170

\bibitem[{{Kamitsukasa} {et~al.}(2015){Kamitsukasa}, {Koyama}, {Uchida},
  {Nakajima}, {Hayashida}, {Mori}, {Katsuda}, \&
  {Tsunemi}}]{2015PASJ...67...16K}
{Kamitsukasa}, F., {Koyama}, K., {Uchida}, H., {et~al.} 2015, \pasj, 67, 16

\bibitem[{{Kamitsukasa} {et~al.}(2014){Kamitsukasa}, {Koyama}, {Tsunemi},
  {Hayashida}, {Nakajima}, {Takahashi}, {Ueda}, {Mori}, {Katsuda}, \&
  {Uchida}}]{2014PASJ...66...64K}
{Kamitsukasa}, F., {Koyama}, K., {Tsunemi}, H., {et~al.} 2014, \pasj, 66, 64

\bibitem[{{Kaspi} {et~al.}(1994){Kaspi}, {Manchester}, {Siegman}, {Johnston},
  \& {Lyne}}]{1994ApJ...422L..83K}
{Kaspi}, V.~M., {Manchester}, R.~N., {Siegman}, B., {Johnston}, S., \& {Lyne},
  A.~G. 1994, \apjl, 422, L83

\bibitem[{{Katsuda} {et~al.}(2009){Katsuda}, {Petre}, {Hwang}, {Yamaguchi},
  {Mori}, \& {Tsunemi}}]{2009PASJ...61S.155K}
{Katsuda}, S., {Petre}, R., {Hwang}, U., {et~al.} 2009, \pasj, 61, S155

\bibitem[{{Katsuda} {et~al.}(2016){Katsuda}, {Tanaka}, {Morokuma}, {Fesen}, \&
  {Milisavljevic}}]{2016ApJ...826..108K}
{Katsuda}, S., {Tanaka}, M., {Morokuma}, T., {Fesen}, R., \& {Milisavljevic},
  D. 2016, \apj, 826, 108

\bibitem[{{Katsuda} {et~al.}(2015){Katsuda}, {Acero}, {Tominaga}, {Fukui},
  {Hiraga}, {Koyama}, {Lee}, {Mori}, {Nagataki}, {Ohira}, {Petre}, {Sano},
  {Takeuchi}, {Tamagawa}, {Tsuji}, {Tsunemi}, \&
  {Uchiyama}}]{2015ApJ...814...29K}
{Katsuda}, S., {Acero}, F., {Tominaga}, N., {et~al.} 2015, \apj, 814, 29

\bibitem[{{Kochanek} {et~al.}(2012){Kochanek}, {Khan}, \&
  {Dai}}]{2012ApJ...759...20K}
{Kochanek}, C.~S., {Khan}, R., \& {Dai}, X. 2012, \apj, 759, 20

\bibitem[{{Koo} {et~al.}(1995){Koo}, {Kim}, \& {Seward}}]{1995ApJ...447..211K}
{Koo}, B.-C., {Kim}, K.-T., \& {Seward}, F.~D. 1995, \apj, 447, 211

\bibitem[{{Kotake} {et~al.}(2012){Kotake}, {Sumiyoshi}, {Yamada}, {Takiwaki},
  {Kuroda}, {Suwa}, \& {Nagakura}}]{2012PTEP.2012aA301K}
{Kotake}, K., {Sumiyoshi}, K., {Yamada}, S., {et~al.} 2012, Progress of
  Theoretical and Experimental Physics, 2012, 01A301

\bibitem[{{Kumar} {et~al.}(2012){Kumar}, {Safi-Harb}, \&
  {Gonzalez}}]{2012ApJ...754...96K}
{Kumar}, H.~S., {Safi-Harb}, S., \& {Gonzalez}, M.~E. 2012, \apj, 754, 96

\bibitem[{{Kumar} {et~al.}(2014){Kumar}, {Safi-Harb}, {Slane}, \&
  {Gotthelf}}]{2014ApJ...781...41K}
{Kumar}, H.~S., {Safi-Harb}, S., {Slane}, P.~O., \& {Gotthelf}, E.~V. 2014,
  \apj, 781, 41

\bibitem[{{Lee} {et~al.}(2014){Lee}, {Park}, {Hughes}, \&
  {Slane}}]{2014ApJ...789....7L}
{Lee}, J.-J., {Park}, S., {Hughes}, J.~P., \& {Slane}, P.~O. 2014, \apj, 789, 7

\bibitem[{{Levenson} {et~al.}(1998){Levenson}, {Graham}, {Keller}, \&
  {Richter}}]{1998ApJS..118..541L}
{Levenson}, N.~A., {Graham}, J.~R., {Keller}, L.~D., \& {Richter}, M.~J. 1998,
  \apjs, 118, 541

\bibitem[{{Lopez} {et~al.}(2011){Lopez}, {Ramirez-Ruiz}, {Huppenkothen},
  {Badenes}, \& {Pooley}}]{2011ApJ...732..114L}
{Lopez}, L.~A., {Ramirez-Ruiz}, E., {Huppenkothen}, D., {Badenes}, C., \&
  {Pooley}, D.~A. 2011, \apj, 732, 114

\bibitem[{{Maeda} {et~al.}(2002){Maeda}, {Baganoff}, {Feigelson}, {Morris},
  {Bautz}, {Brandt}, {Burrows}, {Doty}, {Garmire}, {Pravdo}, {Ricker}, \&
  {Townsley}}]{2002ApJ...570..671M}
{Maeda}, Y., {Baganoff}, F.~K., {Feigelson}, E.~D., {et~al.} 2002, \apj, 570,
  671

\bibitem[{{Maggi} \& {Acero}(2017)}]{2017AA...597A..65M}
{Maggi}, P., \& {Acero}, F. 2017, \aap, 597, A65

\bibitem[{{Maggi} {et~al.}(2016){Maggi}, {Haberl}, {Kavanagh}, {Sasaki},
  {Bozzetto}, {Filipovi{\'c}}, {Vasilopoulos}, {Pietsch}, {Points}, {Chu},
  {Dickel}, {Ehle}, {Williams}, \& {Greiner}}]{2016AA...585A.162M}
{Maggi}, P., {Haberl}, F., {Kavanagh}, P.~J., {et~al.} 2016, \aap, 585, A162

\bibitem[{{Nakamura} {et~al.}(2015){Nakamura}, {Takiwaki}, {Kuroda}, \&
  {Kotake}}]{2015PASJ...67..107N}
{Nakamura}, K., {Takiwaki}, T., {Kuroda}, T., \& {Kotake}, K. 2015, \pasj, 67,
  107

\bibitem[{{Nakano} {et~al.}(2017){Nakano}, {Murakami}, {Furuta}, {Enoto},
  {Masuyama}, {Shigeyama}, \& {Makishima}}]{2017PASJ...69...40N}
{Nakano}, T., {Murakami}, H., {Furuta}, Y., {et~al.} 2017, \pasj, 69, 40

\bibitem[{{Nakano} {et~al.}(2015){Nakano}, {Murakami}, {Makishima}, {Hiraga},
  {Uchiyama}, {Kaneda}, \& {Enoto}}]{2015PASJ...67....9N}
{Nakano}, T., {Murakami}, H., {Makishima}, K., {et~al.} 2015, \pasj, 67, 9

\bibitem[{{Nomoto} {et~al.}(1984){Nomoto}, {Thielemann}, \&
  {Yokoi}}]{1984ApJ...286..644N}
{Nomoto}, K., {Thielemann}, F.-K., \& {Yokoi}, K. 1984, \apj, 286, 644

\bibitem[{{O'Connor} \& {Ott}(2011)}]{2011ApJ...730...70O}
{O'Connor}, E., \& {Ott}, C.~D. 2011, \apj, 730, 70

\bibitem[{{O'Connor} \& {Couch}(2018)}]{2018ApJ...854...63O}
{O'Connor}, E.~P., \& {Couch}, S.~M. 2018, \apj, 854, 63

\bibitem[{{Olbert} {et~al.}(2001){Olbert}, {Clearfield}, {Williams}, {Keohane},
  \& {Frail}}]{2001ApJ...554L.205O}
{Olbert}, C.~M., {Clearfield}, C.~R., {Williams}, N.~E., {Keohane}, J.~W., \&
  {Frail}, D.~A. 2001, \apjl, 554, L205

\bibitem[{{Ott} {et~al.}(2018){Ott}, {Roberts}, {da Silva Schneider}, {Fedrow},
  {Haas}, \& {Schnetter}}]{2018ApJ...855L...3O}
{Ott}, C.~D., {Roberts}, L.~F., {da Silva Schneider}, A., {et~al.} 2018, \apjl,
  855, L3

\bibitem[{{Pannuti} {et~al.}(2010){Pannuti}, {Rho}, {Borkowski}, \&
  {Cameron}}]{2010AJ....140.1787P}
{Pannuti}, T.~G., {Rho}, J., {Borkowski}, K.~J., \& {Cameron}, P.~B. 2010, \aj,
  140, 1787

\bibitem[{{Park} {et~al.}(2003{\natexlab{a}}){Park}, {Hughes}, {Burrows},
  {Slane}, {Nousek}, \& {Garmire}}]{2003ApJ...598L..95P}
{Park}, S., {Hughes}, J.~P., {Burrows}, D.~N., {et~al.} 2003{\natexlab{a}},
  \apjl, 598, L95

\bibitem[{{Park} {et~al.}(2012){Park}, {Hughes}, {Slane}, {Burrows}, {Lee}, \&
  {Mori}}]{2012ApJ...748..117P}
{Park}, S., {Hughes}, J.~P., {Slane}, P.~O., {et~al.} 2012, \apj, 748, 117

\bibitem[{{Park} {et~al.}(2003{\natexlab{b}}){Park}, {Hughes}, {Slane},
  {Burrows}, {Warren}, {Garmire}, \& {Nousek}}]{2003ApJ...592L..41P}
---. 2003{\natexlab{b}}, \apjl, 592, L41

\bibitem[{{Patnaude} {et~al.}(2015){Patnaude}, {Lee}, {Slane}, {Badenes},
  {Heger}, {Ellison}, \& {Nagataki}}]{2015ApJ...803..101P}
{Patnaude}, D.~J., {Lee}, S.-H., {Slane}, P.~O., {et~al.} 2015, \apj, 803, 101

\bibitem[{{Patnaude} {et~al.}(2017){Patnaude}, {Lee}, {Slane}, {Badenes},
  {Nagataki}, {Ellison}, \& {Milisavljevic}}]{2017ApJ...849..109P}
---. 2017, \apj, 849, 109

\bibitem[{{Petre} {et~al.}(1988){Petre}, {Szymkowiak}, {Seward}, \&
  {Willingale}}]{1988ApJ...335..215P}
{Petre}, R., {Szymkowiak}, A.~E., {Seward}, F.~D., \& {Willingale}, R. 1988,
  \apj, 335, 215

\bibitem[{{Rho} {et~al.}(1994){Rho}, {Petre}, {Schlegel}, \&
  {Hester}}]{1994ApJ...430..757R}
{Rho}, J., {Petre}, R., {Schlegel}, E.~M., \& {Hester}, J.~J. 1994, \apj, 430,
  757

\bibitem[{{Rosado}(1983)}]{1983RMxAA...8...59R}
{Rosado}, M. 1983, \rmxaa, 8, 59

\bibitem[{{Sakano} {et~al.}(2004){Sakano}, {Warwick}, {Decourchelle}, \&
  {Predehl}}]{2004MNRAS.350..129S}
{Sakano}, M., {Warwick}, R.~S., {Decourchelle}, A., \& {Predehl}, P. 2004,
  \mnras, 350, 129

\bibitem[{{Salpeter}(1955)}]{1955ApJ...121..161S}
{Salpeter}, E.~E. 1955, \apj, 121, 161

\bibitem[{{Sana} {et~al.}(2012){Sana}, {de Mink}, {de Koter}, {Langer},
  {Evans}, {Gieles}, {Gosset}, {Izzard}, {Le Bouquin}, \&
  {Schneider}}]{2012Sci...337..444S}
{Sana}, H., {de Mink}, S.~E., {de Koter}, A., {et~al.} 2012, Science, 337, 444

\bibitem[{{Sarbadhicary} {et~al.}(2017){Sarbadhicary}, {Badenes}, {Chomiuk},
  {Caprioli}, \& {Huizenga}}]{2017MNRAS.464.2326S}
{Sarbadhicary}, S.~K., {Badenes}, C., {Chomiuk}, L., {Caprioli}, D., \&
  {Huizenga}, D. 2017, \mnras, 464, 2326

\bibitem[{{Sasaki} {et~al.}(2006){Sasaki}, {Gaetz}, {Blair}, {Edgar}, {Morse},
  {Plucinsky}, \& {Smith}}]{2006ApJ...642..260S}
{Sasaki}, M., {Gaetz}, T.~J., {Blair}, W.~P., {et~al.} 2006, \apj, 642, 260

\bibitem[{{Sasaki} {et~al.}(2014){Sasaki}, {Heinitz}, {Warth}, \&
  {P{\"u}hlhofer}}]{2014AA...563A...9S}
{Sasaki}, M., {Heinitz}, C., {Warth}, G., \& {P{\"u}hlhofer}, G. 2014, \aap,
  563, A9

\bibitem[{{Sato} {et~al.}(2016){Sato}, {Koyama}, {Lee}, \&
  {Takahashi}}]{2016PASJ...68S...8S}
{Sato}, T., {Koyama}, K., {Lee}, S.-H., \& {Takahashi}, T. 2016, \pasj, 68, S8

\bibitem[{{Sato} {et~al.}(2014){Sato}, {Koyama}, {Takahashi}, {Odaka}, \&
  {Nakashima}}]{2014PASJ...66..124S}
{Sato}, T., {Koyama}, K., {Takahashi}, T., {Odaka}, H., \& {Nakashima}, S.
  2014, \pasj, 66, 124

\bibitem[{{Slane} {et~al.}(2002{\natexlab{a}}){Slane}, {Chen}, {Lazendic}, \&
  {Hughes}}]{2002ApJ...580..904S}
{Slane}, P., {Chen}, Y., {Lazendic}, J.~S., \& {Hughes}, J.~P.
  2002{\natexlab{a}}, \apj, 580, 904

\bibitem[{{Slane} {et~al.}(2002{\natexlab{b}}){Slane}, {Smith}, {Hughes}, \&
  {Petre}}]{2002ApJ...564..284S}
{Slane}, P., {Smith}, R.~K., {Hughes}, J.~P., \& {Petre}, R.
  2002{\natexlab{b}}, \apj, 564, 284

\bibitem[{{Smartt}(2015)}]{2015PASA...32...16S}
{Smartt}, S.~J. 2015, \pasa, 32, e016

\bibitem[{{Smartt} {et~al.}(2009){Smartt}, {Eldridge}, {Crockett}, \&
  {Maund}}]{2009MNRAS.395.1409S}
{Smartt}, S.~J., {Eldridge}, J.~J., {Crockett}, R.~M., \& {Maund}, J.~R. 2009,
  \mnras, 395, 1409

\bibitem[{{Su} {et~al.}(2011){Su}, {Chen}, {Yang}, {Koo}, {Zhou}, {Lu},
  {Jeong}, \& {DeLaney}}]{2011ApJ...727...43S}
{Su}, Y., {Chen}, Y., {Yang}, J., {et~al.} 2011, \apj, 727, 43

\bibitem[{{Sukhbold} {et~al.}(2016){Sukhbold}, {Ertl}, {Woosley}, {Brown}, \&
  {Janka}}]{2016ApJ...821...38S}
{Sukhbold}, T., {Ertl}, T., {Woosley}, S.~E., {Brown}, J.~M., \& {Janka}, H.-T.
  2016, \apj, 821, 38

\bibitem[{{Summa} {et~al.}(2016){Summa}, {Hanke}, {Janka}, {Melson}, {Marek},
  \& {M{\"u}ller}}]{2016ApJ...825....6S}
{Summa}, A., {Hanke}, F., {Janka}, H.-T., {et~al.} 2016, \apj, 825, 6

\bibitem[{{Suzuki} \& {Maeda}(2018)}]{2018ApJ...852..101S}
{Suzuki}, A., \& {Maeda}, K. 2018, \apj, 852, 101

\bibitem[{{Takeuchi} {et~al.}(2016){Takeuchi}, {Yamaguchi}, \&
  {Tamagawa}}]{2016PASJ...68S...9T}
{Takeuchi}, Y., {Yamaguchi}, H., \& {Tamagawa}, T. 2016, \pasj, 68, S9

\bibitem[{{Takiwaki} {et~al.}(2016){Takiwaki}, {Kotake}, \&
  {Suwa}}]{2016MNRAS.461L.112T}
{Takiwaki}, T., {Kotake}, K., \& {Suwa}, Y. 2016, \mnras, 461, L112

\bibitem[{{Temim} {et~al.}(2017){Temim}, {Slane}, {Plucinsky}, {Gelfand},
  {Castro}, \& {Kolb}}]{2017ApJ...851..128T}
{Temim}, T., {Slane}, P., {Plucinsky}, P.~P., {et~al.} 2017, \apj, 851, 128

\bibitem[{{Tian} \& {Leahy}(2008)}]{2008ApJ...677..292T}
{Tian}, W.~W., \& {Leahy}, D.~A. 2008, \apj, 677, 292

\bibitem[{{Tominaga} {et~al.}(2007){Tominaga}, {Umeda}, \&
  {Nomoto}}]{2007ApJ...660..516T}
{Tominaga}, N., {Umeda}, H., \& {Nomoto}, K. 2007, \apj, 660, 516

\bibitem[{{Troja} {et~al.}(2008){Troja}, {Bocchino}, {Miceli}, \&
  {Reale}}]{2008AA...485..777T}
{Troja}, E., {Bocchino}, F., {Miceli}, M., \& {Reale}, F. 2008, \aap, 485, 777

\bibitem[{{Tsunemi} {et~al.}(2007){Tsunemi}, {Katsuda}, {Nemes}, \&
  {Miller}}]{2007ApJ...671.1717T}
{Tsunemi}, H., {Katsuda}, S., {Nemes}, N., \& {Miller}, E.~D. 2007, \apj, 671,
  1717

\bibitem[{{Uchida} {et~al.}(2015){Uchida}, {Koyama}, \&
  {Yamaguchi}}]{2015ApJ...808...77U}
{Uchida}, H., {Koyama}, K., \& {Yamaguchi}, H. 2015, \apj, 808, 77

\bibitem[{{Uchida} {et~al.}(2009){Uchida}, {Tsunemi}, {Katsuda}, {Kimura}, \&
  {Kosugi}}]{2009PASJ...61..301U}
{Uchida}, H., {Tsunemi}, H., {Katsuda}, S., {Kimura}, M., \& {Kosugi}, H. 2009,
  \pasj, 61, 301

\bibitem[{{Uchida} {et~al.}(2012{\natexlab{a}}){Uchida}, {Tsunemi}, {Katsuda},
  {Mori}, {Petre}, \& {Yamaguchi}}]{2012PASJ...64...61U}
{Uchida}, H., {Tsunemi}, H., {Katsuda}, S., {et~al.} 2012{\natexlab{a}}, \pasj,
  64, 61

\bibitem[{{Uchida} {et~al.}(2012{\natexlab{b}}){Uchida}, {Koyama}, {Yamaguchi},
  {Sawada}, {Ohnishi}, {Tsuru}, {Tanaka}, {Yoshiike}, \&
  {Fukui}}]{2012PASJ...64..141U}
{Uchida}, H., {Koyama}, K., {Yamaguchi}, H., {et~al.} 2012{\natexlab{b}},
  \pasj, 64, 141

\bibitem[{{Ugliano} {et~al.}(2012){Ugliano}, {Janka}, {Marek}, \&
  {Arcones}}]{2012ApJ...757...69U}
{Ugliano}, M., {Janka}, H.-T., {Marek}, A., \& {Arcones}, A. 2012, \apj, 757,
  69

\bibitem[{{van der Heyden} {et~al.}(2004){van der Heyden}, {Bleeker}, \&
  {Kaastra}}]{2004AA...421.1031V}
{van der Heyden}, K.~J., {Bleeker}, J.~A.~M., \& {Kaastra}, J.~S. 2004, \aap,
  421, 1031

\bibitem[{{Van Dyk}(2017)}]{2017RSPTA.37560277V}
{Van Dyk}, S.~D. 2017, Philosophical Transactions of the Royal Society of
  London Series A, 375, 20160277

\bibitem[{{Vartanyan} {et~al.}(2018){Vartanyan}, {Burrows}, {Radice},
  {Skinner}, \& {Dolence}}]{2018MNRAS.tmp..786V}
{Vartanyan}, D., {Burrows}, A., {Radice}, D., {Skinner}, M.~A., \& {Dolence},
  J. 2018, \mnras, arXiv:1801.08148

\bibitem[{{Vink}(2012)}]{2012AARv..20...49V}
{Vink}, J. 2012, \aapr, 20, 49

\bibitem[{{Vogt} \& {Dopita}(2011)}]{2011ApSS.331..521V}
{Vogt}, F., \& {Dopita}, M.~A. 2011, \apss, 331, 521

\bibitem[{{Walmswell} \& {Eldridge}(2012)}]{2012MNRAS.419.2054W}
{Walmswell}, J.~J., \& {Eldridge}, J.~J. 2012, \mnras, 419, 2054

\bibitem[{{Wang} {et~al.}(1997){Wang}, {Qu}, \& {Chen}}]{1997AA...318L..59W}
{Wang}, Z.~R., {Qu}, Q.-Y., \& {Chen}, Y. 1997, \aap, 318, L59

\bibitem[{{Washino} {et~al.}(2016){Washino}, {Uchida}, {Nobukawa}, {Tsuru},
  {Tanaka}, {Kawabata Nobukawa}, \& {Koyama}}]{2016PASJ...68S...4W}
{Washino}, R., {Uchida}, H., {Nobukawa}, M., {et~al.} 2016, \pasj, 68, S4

\bibitem[{{Weltevrede} {et~al.}(2011){Weltevrede}, {Johnston}, \&
  {Espinoza}}]{2011MNRAS.411.1917W}
{Weltevrede}, P., {Johnston}, S., \& {Espinoza}, C.~M. 2011, \mnras, 411, 1917

\bibitem[{{Williams} {et~al.}(2018){Williams}, {Hillis}, {Murphy}, {Gilbert},
  {Dalcanton}, \& {Dolphin}}]{2018arXiv180308112W}
{Williams}, B.~F., {Hillis}, T.~J., {Murphy}, J.~W., {et~al.} 2018, ArXiv
  e-prints, arXiv:1803.08112

\bibitem[{{Williams} {et~al.}(2015){Williams}, {Rangelov}, {Kargaltsev}, \&
  {Pavlov}}]{2015ApJ...808L..19W}
{Williams}, B.~J., {Rangelov}, B., {Kargaltsev}, O., \& {Pavlov}, G.~G. 2015,
  \apjl, 808, L19

\bibitem[{{Winkler} {et~al.}(2009){Winkler}, {Twelker}, {Reith}, \&
  {Long}}]{2009ApJ...692.1489W}
{Winkler}, P.~F., {Twelker}, K., {Reith}, C.~N., \& {Long}, K.~S. 2009, \apj,
  692, 1489

\bibitem[{{Wolszczan} {et~al.}(1991){Wolszczan}, {Cordes}, \&
  {Dewey}}]{1991ApJ...372L..99W}
{Wolszczan}, A., {Cordes}, J.~M., \& {Dewey}, R.~J. 1991, \apjl, 372, L99

\bibitem[{{Woosley} \& {Weaver}(1995)}]{1995ApJS..101..181W}
{Woosley}, S.~E., \& {Weaver}, T.~A. 1995, \apjs, 101, 181

\bibitem[{{Yamaguchi} {et~al.}(2009){Yamaguchi}, {Bamba}, \&
  {Koyama}}]{2009PASJ...61S.175Y}
{Yamaguchi}, H., {Bamba}, A., \& {Koyama}, K. 2009, \pasj, 61, S175

\bibitem[{{Yamaguchi} {et~al.}(2014){Yamaguchi}, {Badenes}, {Petre}, {Nakano},
  {Castro}, {Enoto}, {Hiraga}, {Hughes}, {Maeda}, {Nobukawa}, {Safi-Harb},
  {Slane}, {Smith}, \& {Uchida}}]{2014ApJ...785L..27Y}
{Yamaguchi}, H., {Badenes}, C., {Petre}, R., {et~al.} 2014, \apjl, 785, L27

\bibitem[{{Yasumi} {et~al.}(2014){Yasumi}, {Nobukawa}, {Nakashima}, {Uchida},
  {Sugawara}, {Tsuru}, {Tanaka}, \& {Koyama}}]{2014PASJ...66...68Y}
{Yasumi}, M., {Nobukawa}, M., {Nakashima}, S., {et~al.} 2014, \pasj, 66, 68

\bibitem[{{Yatsu} {et~al.}(2013){Yatsu}, {Asano}, {Kawai}, {Yano}, \&
  {Nakamori}}]{2013ApJ...773...25Y}
{Yatsu}, Y., {Asano}, K., {Kawai}, N., {Yano}, Y., \& {Nakamori}, T. 2013,
  \apj, 773, 25

\bibitem[{{Yatsu} {et~al.}(2005){Yatsu}, {Kawai}, {Kataoka}, {Kotani},
  {Tamura}, \& {Brinkmann}}]{2005ApJ...631..312Y}
{Yatsu}, Y., {Kawai}, N., {Kataoka}, J., {et~al.} 2005, \apj, 631, 312

\bibitem[{{Yoon} {et~al.}(2012){Yoon}, {Gr{\"a}fener}, {Vink}, {Kozyreva}, \&
  {Izzard}}]{2012AA...544L..11Y}
{Yoon}, S.-C., {Gr{\"a}fener}, G., {Vink}, J.~S., {Kozyreva}, A., \& {Izzard},
  R.~G. 2012, \aap, 544, L11

\bibitem[{{Zhou} {et~al.}(2016){Zhou}, {Chen}, {Safi-Harb}, {Zhou}, {Sun},
  {Zhang}, \& {Zhang}}]{2016ApJ...831..192Z}
{Zhou}, P., {Chen}, Y., {Safi-Harb}, S., {et~al.} 2016, \apj, 831, 192

\end{thebibliography}

%\begin{thebibliography}{}
%\expandafter\ifx\csname natexlab\endcsname\relax\def\natexlab#1{#1}\fi
%
%\bibitem[{{Arnaud}(1996)}]{1996ASPC..101...17A}
%{Arnaud}, K.~A. 1996, in Astronomical Society of the Pacific Conference Series,
%  Vol. 101, Astronomical Data Analysis Software and Systems V, ed. G.~H.
%  {Jacoby} \& J.~{Barnes}, 17
%
%
%\end{thebibliography}

\newpage

%\floattable
\begin{deluxetable}{ccccccc}
%\tablenum{1}
\tabletypesize{\tiny}
\tablecaption{Abundance-based progenitor masses of core-collapse SNRs in our Galaxy and Magellanic Clouds\label{tab:snrs}}
\tablewidth{0pt}
\tablehead{
\colhead{SNR} & Age (yr) & \colhead{M$_{\rm ZAMS}$ (M$_\odot$)} & \colhead{(Fe/Si)/(Fe/Si)$_\odot$} & \colhead{Revised M$_{\rm COcore}$ (M$_\odot$)} & \colhead{Revised M$_{\rm ZAMS}$ (M$_\odot$)}
}
%\decimalcolnumbers
\startdata
*** Galactic SNRs *** & & & & & \\
Cassiopeia~A & $\sim$340 [1] & 15--20 [2] & 1.0$\pm$0.1 [3] & $<$ 3 & $<$ 15 \\
Kes~73 & $\sim$750 [4] & 20--30 [5] & 0.8$^{+1.0}_{-0.3}$ [5] & $<$ 3 & $<$ 15$^{\rm m1}$ \\
G350.1-0.3 & $\sim$900 [6] & 15--25 [7] & 0.35$\pm$0.05 [7] & 3--6 & 15--22.5$^{\rm m1}$ \\
RX~J1713.7-3946 & $\sim$1600 [8] & $\lesssim$15 [9] & $<$0.03 [9] & $>$ 6 & $>$ 22.5 \\
MSH 15-52 & $\sim$1700 [10] & N.A. & 0.78$\pm$0.09 [11] & $<$ 3 & $<$ 15 \\
G292.2-0.5 & $\sim$1900 [12] & 25--30 [13] & 0.59$^{+0.62}_{-0.48}$ [13] & $<$ 3 & $<$ 15$^{\rm m1,m2}$ \\
RCW103 & $\sim$2000 [14] & 18--20 [15] & 1.33$^{+0.27}_{-0.14}$ [15] & $<$ 3 & $<$ 15 \\
G349.7+0.2 & $\sim$2800 [16] & 35--40 [7] & 0.56$^{+0.09}_{-0.10}$ [7] & $<$ 3 & $<$ 15$^{\rm m1}$\\
G292.0+1.8 & $\sim$3000 [17] & 30--35 [18] & 0.55$\pm$0.24 [18] & $<$ 3 & $<$ 15$^{\rm m1,m2}$ \\
Puppis~A & $\sim$4500 & 15--25 [19] & 0.63$\pm$0.05 [20] & $<$ 3 & $<$ 15  \\
Kes~79 & 4400--6700 [21] & 30--40 [22] & 0.35$^{+0.04}_{-0.05}$ [22] & 3--6 & 15--22.5$^{\rm m1}$ \\
Cygnus Loop & $\sim$10000 [23] & $\lesssim$15 [24] & 0.7$\pm$0.1 [25] & $<$ 3 & $<$ 15 \\
Sgr A East & $\sim$10000 [26] & 13--20 [26] & 0.26$^{+0.12}_{-0.09}$ [27] & 3--6 & 15--22.5$^{\rm m1,m2}$ \\
MSH 15-56 & $\sim$11000 [28] & N.A. & 0.37$\pm$0.11 [29] & 3--6 & 15--22.5$^{\rm m1}$ \\
IC443 & 3000--30000 [30,31] & $\sim$25 (32) & 0.25$\pm$0.10 (33) & 3--6 & 15--22.5$^{\rm m1,m2}$ \\
G290.1-0.8 & 10000--20000 [34] & 20--25 [35] & 0.11$\pm$0.06 [35] & $>$ 6 & $>$ 22.5 \\
3C391 & $\sim$19000 [36] & $\sim$15 [37] & $<$0.06 [37] & $>$ 6 & $>$ 22.5 \\
W44 & 20000 [38] & 8--15 [39] & 0.03$\pm$0.01 [40] & $>$ 6 & $>$ 22.5$^{\rm m2}$ \\
G284.3-1.8 & $\sim$21000 [41] & $>$ 25 [42] & 0.59$^{+1.59}_{-0.56}$ [42] & $<$ 3 & $<$ 15$^{\rm m1,m2}$ \\ 
G156.2+5.7 & 20000--30000 [43] & $\lesssim$15 [44] & 0.37$\pm$0.1 [44,45] & 3--6 & 15--22.5$^{\rm m1}$ \\
3C400.2 & $\sim$100000 [46] & N.A. & 5.3$^{+3.1}_{-2.1}$ [47] & $<$ 3 & $<$ 15 \\
3C396 & $\sim$3000 [48] & 13--15 [48] & N.A. & N.A. & N.A. \\
G15.9+0.2 & 2000--6000 [49] & 20--25 [49] & N.A. & N.A. & N.A. \\
Kes~17 & 2000--40000 [50] & 25--30 [51] & N.A. & N.A. & N.A.  \\
CTB109 & $\sim$14000 [52] & 30--40 [53] & N.A. & N.A. & N.A. \\
G116.9+0.2 (CTB1) & $\sim$16000 [54] & 13--15 [55] & N.A. & N.A. & N.A. \\
G296.1-0.5 & $\sim$28000 [56] & 25--30 [56] & N.A. & N.A. & N.A. \\
W51C & $\sim$30000 [57] & $\gtrsim$20 [58] & N.A. & N.A. & N.A. \\
\hline
*** LMC SNRs *** & & & & &\\
N132D & $\sim$2500 [59] & $\sim$50 [60] & 0.48$^{+0.14}_{-0.25}$ [61] & $<$ 3 & $<$ 15$^{\rm m1}$ \\
N63A & 2000--5000 [62] & N.A. & 0.87$\pm$0.13 [63] & $<$ 3 & $<$ 15 \\
N23 & $\sim$4000 [64] & N.A. & 0.38$\pm$0.13 [65] & 3--6 & 15--22.5$^{\rm m1}$ \\
N49 & $\sim$4800 [66] & N.A. & 0.18$\pm$0.01 [65] & $>$ 6 & $>$ 22.5$^{\rm m2}$ \\
N49B & $\sim$10000 [62] & $>$25 [64] & 1.03$\pm$0.07 [65] & $<$ 3 & $<$ 15 \\
B0453-68.5 & 12000--15000 [68] & N.A. & 0.42$^{+0.17}_{-0.16}$ [63] & 3--6 & 15--22.5$^{\rm m1,m2}$\\
30 Dor C & 4000--20000 [69] & N.A. & 0.08$^{+0.20}_{-0.06}$ [69] & $>$ 6 & $>$ 22.5$^{\rm m2}$ \\
Honeycomb & N.A. & N.A. & 0.17$^{+0.13}_{-0.10}$ [63] & $>$ 6 & $>$ 22.5$^{\rm m2}$  \\
%SNR~1987A & 30 & (20) & 0.39$^{+0.02}_{-0.01}$ & 3--6 &  15--22.5$^{\rm m1}$\\
\hline
*** SMC SNRs *** & & & & &\\
1E0102.2-7219 & $\sim$2050 [70] & 25--35 [71] & 0.63$^{+0.26}_{-0.20}$ [72] & $<$ 3 & $<$ 15$^{\rm m1}$ \\
IKT2 & N.A. & N.A. & 0.32$\pm$0.24 [73] & 3--6 & 15--22.5$^{\rm m1,m2}$ \\
DEM S32 & N.A. & N.A. & 0.28$\pm$0.26 [73] & 3--6 & 15--22.5$^{\rm m1,m2}$\\
IKT6 & $\sim$14000 [74] & N.A. & 0.26$^{+0.16}_{-0.07}$ [75] & 3--6 & 15--22.5$^{\rm m1,m2}$  \\
IKT23 & $\sim$18000 [76] & $\sim$18 [76] & 0.48$^{+0.14}_{-0.25}$ [75] & $<$ 3 & $<$ 15$^{\rm m1,m2}$ \\
\enddata
\tablecomments{$^a$ The marks m1 and m2 in parentheses represent marginal data at M$_{\rm ZAMS} =$ 15\,M$_\odot$ and M$_{\rm ZAMS} =$ 22.5\,M$_\odot$, respectively.  
References: 1. \citet{2006ApJ...645..283F}, 2. \citet{2014ApJ...789....7L}, 3. \citet{2012ApJ...746..130H},
4. \citet{2008ApJ...677..292T}, 5. \citet{2014ApJ...781...41K},
6. \citet{2008ApJ...680L..37G}, 7. \citet{2014PASJ...66...68Y}, 
8. \citet{1997AA...318L..59W}, 9. \citet{2015ApJ...814...29K},
10. \citet{1994ApJ...422L..83K}, 11. \citet{2005ApJ...631..312Y},
12. \citet{2011MNRAS.411.1917W}, 13. \citet{2012ApJ...754...96K}, 
14. \citet{1997PASP..109..990C}, 15. \citet{2015ApJ...810..113F},
16. \citet{2002ApJ...580..904S}, 
17. \citet{2009ApJ...692.1489W}, 18. \citet{2014PASJ...66...64K},
19. \citet{2012ApJ...755..141B}, 20. \citet{2008ApJ...676..378H},
21. \citet{2016ApJ...831..192Z}, 22. \citet{2016PASJ...68S...8S},
23. \citet{1998ApJS..118..541L}, 24. \citet{2007ApJ...671.1717T}, 25. \citet{2009PASJ...61..301U},
26. \citet{2002ApJ...570..671M}, 27. \citet{2004MNRAS.350..129S},
28. \citet{2017ApJ...851..128T}, 29. \citet{2013ApJ...773...25Y},
30. \citet{1988ApJ...335..215P}, 31. \citet{2001ApJ...554L.205O}, 32. \citet{2008AA...485..777T}, 33. \citet{2009AA...498..139B},
34. \citet{2002ApJ...564..284S}, 35. \citet{2015PASJ...67...16K}, 
36. \citet{2001ApJ...563..202C}, 37. \citet{2014PASJ...66..124S}, 
38. \citet{1991ApJ...372L..99W}, 39. \citet{1994ApJ...430..757R}, 40. \citet{2012PASJ...64..141U},
41. \citet{2004ApJ...616.1118C}, 42. \citet{2015ApJ...808L..19W},
43. \citet{2016ApJ...826..108K}, 44. \citet{2009PASJ...61S.155K}, 45. \citet{2012PASJ...64...61U},
46. \citet{1983RMxAA...8...59R}, 47. \citet{2015MNRAS.446.3885B},
48. \citet{2011ApJ...727...43S},
49. \citet{2017AA...597A..65M}, 
50. \citet{2013ApJ...777..148G}, 51. \citet{2016PASJ...68S...4W},
52. \citet{2015PASJ...67....9N}, 53. \citet{2017PASJ...69...40N}, 
54. \citet{1997AJ....113..767F}, 55. \citet{2010AJ....140.1787P},
56. \citet{2012MNRAS.419.1603G}, 
57. \citet{1995ApJ...447..211K}, 58. \citet{2014AA...563A...9S}, 
59. \citet{2011ApSS.331..521V}, 60. \citet{2009ApJ...707L..27F}, 61. \citet{2018ApJ...854...71B},
62. \citet{1998ApJ...505..732H}, 63. \citet{2016AA...585A.162M},
64. \citet{2011AA...535A..11B}, 65. \citet{2015ApJ...808...77U},
66. \citet{2012ApJ...748..117P}, 
67. \citet{2003ApJ...592L..41P},
68. \citet{2012AA...543A.154H},
69. \citet{2009PASJ...61S.175Y}, 
70. \citet{2006ApJ...641..919F}, 71. \citet{2000ApJ...537..667B}, 72. \citet{2006ApJ...642..260S}, 
73. \citet{2004AA...421.1031V}, 
74. \citet{2005ApJ...622L.117H}, 75. \citet{2016PASJ...68S...9T},
76. \citet{2003ApJ...598L..95P}.
}
\end{deluxetable}
\clearpage

%\floattable
\begin{deluxetable}{lcccccc}
%\tablenum{1}
\tabletypesize{\tiny}
\tablecaption{Progenitor mass distributions \label{tab:frac}}
\tablewidth{0pt}
\tablehead{
\colhead{Sample} & \colhead{$f$(M$_{\rm ZAMS} < 15$\,M$_\odot$)} & \colhead{$f$(15\,M$_\odot <$ M$_{\rm ZAMS} < 22.5$\,M$_\odot$)} & \colhead{$f$(22.5\,M$_\odot$ $<$ M$_{\rm ZAMS}$)} 
}
%\decimalcolnumbers
\startdata
Data (literature) & 0.24$^{+0.03}_{-0.17}$ & 0.28$^{+0.21}_{-0.17}$ & 0.48$^{+0.14}_{-0.07}$ \\
Data (this work including all SNRs) & 0.47$^{+0.35}_{-0.24}$ & 0.32$\pm$0.32 & 0.21$^{+0.26}_{-0.12}$ \\
Data (this work restricted to young SNRs with $t \lesssim 5000$\,yr) & 0.69$^{+0.19}_{-0.38}$ & 0.19$^{+0.50}_{-0.19}$ & 0.12$^{+0.12}_{-0.06}$ \\
IMF ($\alpha$=-2.35, M$_{\rm min}$=8\,M$_\odot$, M$_{\rm max}$=40\,M$_\odot$)$^a$ & 0.65 & 0.20 & 0.15 \\ 
IMF ($\alpha$=-3.0, M$_{\rm min}$=8\,M$_\odot$, M$_{\rm max}$=40\,M$_\odot$)$^b$ & 0.75 & 0.16 & 0.09 \\ 
IMF ($\alpha$=-2.35, M$_{\rm min}$=8\,M$_\odot$, M$_{\rm max}$=16.5\,M$_\odot$)$^c$ & 0.91 & 0.09 & 0.0 
\enddata
\tablecomments{$^a$The Salpeter IMF.  $^b$Progenitor mass distribution for CCSNRs \citep{2012ApJ...761...26J,2014ApJ...795..170J,2018arXiv180207870D}.  $^c$Progenitor mass distribution for SNe IIP \citep{2009MNRAS.395.1409S,2015PASA...32...16S}.}
\end{deluxetable}

\end{document}